\documentclass[useAMS,usenatbib]{mn2e}

\def\Msun {\,\mathrm{M}_\odot}
\def\CII {C{\sc ii}\,}

\usepackage{amsmath}
\usepackage{amssymb}
\usepackage{graphicx}
\usepackage{color}
\usepackage{url}
\title[AGN outflows in galaxy discs]{Active galactic nucleus outflows in galaxy discs}

\author[T. Hartwig, M. Volonteri, and G. Dashyan]
{\parbox{\textwidth}{
Tilman Hartwig\thanks{E-mail: tilman.hartwig@ipmu.jp}, Marta Volonteri, Gohar Dashyan\\
}\\
Sorbonne Universit\'es, UPMC Univ Paris 06, CNRS \& UMR 7095, Institut d'Astrophysique de Paris, F-75014 Paris, France
}

\begin{document}


\pagerange{\pageref{firstpage}--\pageref{lastpage}} \pubyear{2018}

\maketitle

\label{firstpage}

\begin{abstract}
Galactic outflows, driven by active galactic nuclei (AGNs), play a crucial role in galaxy formation and in the self-regulated growth of supermassive black holes (BHs).
AGN feedback couples to and affects gas, rather than stars, and in many, if not most, gas-rich galaxies cold gas is rotationally supported and settles in a disc. We present a 2D analytical model for AGN-driven outflows in a gaseous disc and demonstrate the main improvements, compared to existing 1D solutions.
We find significant differences for the outflow dynamics and wind efficiency. The outflow is energy-driven due to inefficient cooling up to a certain AGN luminosity ($\sim 10^{43}$\,erg\,s$^{-1}$ in our fiducial model), above which the outflow remains momentum-driven in the disc up to galactic scales. We reproduce results of 3D simulations that gas is preferentially ejected perpendicular to the disc and find that the fraction of ejected interstellar medium is lower than in 1D models. The recovery time of gas in the disc, defined as the freefall time from the radius to which the AGN pushes the ISM at most, is remarkably short, of the order 1\,Myr. This indicates that AGN-driven winds cannot suppress BH growth for long.
Without the inclusion of supernova feedback, we find a scaling of the BH mass with the halo velocity dispersion of $M_\mathrm{BH} \propto \sigma ^{4.8}$.
\end{abstract}

\begin{keywords}
black hole physics -- galaxies: active -- quasars: general -- quasars: supermassive black holes -- methods: analytical
\end{keywords}

\section{Introduction}
The majority of galaxies in the local and distant universe host supermassive black holes (SMBHs) in their centres. The accretion discs around SMBHs are nature's most efficient engines to convert gravitational energy of the infalling gas into radiation.
These extreme sources of energy are expected to have a strong feedback effect on the galactic gas content and star formation. AGN feedback may also self-regulate the growth of the central SMBH and its inclusion in models of galaxy formation improves the match between the simulated and the observed galaxy luminosity function for massive galaxies \citep{bower06,croton06,somerville08,hirschmann14,croton16}.

There are several observations that indicate a correlation between the mass of the central SMBH and large-scale properties of the host galaxy, such as the stellar velocity dispersion $\sigma$, luminosity, or the bulge mass \citep{maborrian98,ferrarese00,gebhardt00,tremaine02,marconi03,haering04,gueltekin09}. These relations imply a possible co-evolution of the black hole (BH) and its host galaxy, and AGN feedback has been suggested to be responsible for regulating  accretion on the BH and star formation in the galaxy, possibly guiding this correlation \citep[see][for a review]{heckman11}. The co-evolution, however, is still debated and there are many open questions regarding the nature of AGN feedback (recent reviews on the observational evidence and the theoretical modelling of AGN feedback can be found in \citealt{fabian12,king15}). Is this relation between the BH mass and galactic properties still valid in galaxies at higher redshift or with lower masses \citep{greene08,baldassare15,reines15,habouzit17}? How and in which form is the energy from the AGN on sub-pc scales communicated to the interstellar medium (ISM) on kpc scales? Under which conditions can gas escape the gravitational potential of a galaxy? How could the first BHs grow to masses above $10^9\Msun$ within the first billion years of the universe, although strong radiative feedback is expected to suppress further gas inflow on to the BH?


Our current understanding of AGN-driven outflows is based on observations at various wavelengths. The dynamics of the ISM, which is a first indicator of AGN feedback, can be traced by dust, CO, C$^+$, or C{\sc ii} emission with radio telescopes, such as ALMA. The observations of gas kinematics of high-z galaxies is one of ALMA's main science drivers and it can reach sub-kpc resolution at $z = 6$ \citep{wang13,t17,decarli18}.
Radio observations reveal large molecular outflows with velocities of $100\,\mathrm{km}\,\mathrm{s}^{-1}$ up to a few times $1000\,\mathrm{km}\,\mathrm{s}^{-1}$ \citep{aalto12b,cicone14}. Recently, \citet{maiolino17} report star formation inside a galactic outflow, another strong indication for the presence of an outflow of cold gas. In Mrk231, the closest and best-studied quasar, we observe velocities of $\sim 1000\,\mathrm{km}\,\mathrm{s}^{-1}$ and mass outflow rates of the order $\sim 1000\Msun\,\mathrm{yr}^{-1}$ \citep{feruglio10,rupke11,sturm11}. These outflows have to be powered by the AGN, because SN-driven outflows can only account for outflow velocities up to $\sim 600\,\mathrm{km}\,\mathrm{s}^{-1}$ \citep{martin05,sharma13}. Besides these molecular outflows, the Fe K lines in the X-ray reveal highly ionized AGN outflows with mildly relativistic velocities of $\sim 0.1c$ \citep{chartas02,pounds03,reeves03,cappi06,gofford13}. Some ultrafast outflows even have velocities up to $\sim 0.3 c$, but the majority has velocities around $\sim 0.1c$ \citep{king15}.

The consensual picture of AGN-driven outflows consists of an inner, line-driven disc wind with mildly relativistic velocities (also called ``nuclear wind''). This highly ionized wind is shocked once it encounters the denser ISM and drives a forward shock into the ISM, which sweeps up most of the gas. As we  discuss below, the details of the driving mechanism depend on the efficiency of  cooling in the shocked wind: if the internal energy is radiated away, only the momentum input from the AGN drives the outflow (``momentum-driven''). If cooling is inefficient, the hot shocked wind adiabatically expands and drives the shock into the ISM (``energy-driven''). These different driving mechanisms have fundamental influence on the outflow dynamics and the thermal state of the outflow. Most of the thermal energy is carried by the shocked wind and the shocked ISM accounts for the bulk part of the kinetic energy.

Several groups have developed 1D analytical models of AGN-driven outflows in these two different regimes \citep{silk98,fabian99,king03,murray05,silk10,ishibashi12,zubovas12,FG12}. While these models can reproduce several observations, the assumption of a spherically symmetric gas distribution is too simplistic to capture the complexity of a realistic galaxy. Moreover, these models often assume an energy- or momentum-driven outflow, without self-consistently solving for the corresponding transition. For a momentum-driven outflow, you find a characteristic BH mass of \citep{king15}
\begin{equation}
M_\sigma  = \frac{f_\mathrm{gas}\kappa}{\pi G} \sigma ^4,
\end{equation}
with the gas opacity $\kappa$, above which the AGN can eject the gas out of the gravitational potential of its host galaxy and hence prevents further BH growth. The $\propto \sigma ^4$ scaling is independent of the underlying density distribution \citep{McQuillin12}. If, on the other hand, we assume an energy-driven outflow, this characteristic mass and its scaling with the velocity dispersion is given by
\begin{equation}
M_\sigma = \frac{11 f_\mathrm{gas} \kappa}{\eta \pi G^2 c} \sigma ^5,
\end{equation}
where $\eta$ is proportional to the mechanical luminosity \citep{silk98,haehnelt98,fabian12,McQuillin13}. Observationally, different values for the exponent $\alpha$ of the M-$\sigma$ relation, $M_\mathrm{BH} \propto \sigma ^\alpha$, are proposed, ranging from $\alpha = 3.7$ \citep{gebhardt00} to $\alpha=5.6$ \citep{McC13} with other values within this range \citep[see e.g.][]{ferrarese00,tremaine02,gueltekin09,kormendy13}.

Numerical simulations of AGN-driven outflows in 3D demonstrate that the outflow takes the path of least resistance and propagates preferentially towards the steepest density gradient, i.e. along the poles of the host galaxy \citep{gaspari11,gaspari13,gabor13,gabor14,costa14b,costa15,roos15,bieri17,barai17}. This results in highly anisotropic outflows as found by \citet{costa14} and observationally confirmed by \citet{cicone15}. However, 3D simulations remain expensive, have explored limited parameter space, and have not yet focused on low-mass or high-redshift galaxies.

Our aim is to develop a flexible, but general model to study AGN feedback in galaxies where rotational support in the gas component is non-negligible. We derive a 2D analytical model for AGN-driven outflows: two dimensions are sufficient to capture the geometry of a disc-like galaxy, which allows simultaneously an AGN-feeding inflow and an ISM-ejecting outflow, but it is still simple enough to be treated analytically and flexible enough to explore a large parameter space with this model. We include a self-consistent transition from momentum- to energy-driving and derive observables, such as velocities, mass outflow rates, and momentum loading.

\section{Methodology}
In this section we describe the basic equations that govern our 2D analytical approach, justify necessary approximations, and introduce our physically motivated galaxy model.

\subsection{Galaxy model}
We present and motivate the individual components and characteristic radii of our analytical approach.

\subsubsection{Gas density profile of the disc}
Our main focus is to model AGN-driven outflows in low-mass galaxies, which represent the first galaxies at high redshift or dwarf galaxies at lower redshift \citep{penny17}. We assume that the gas distribution in these galaxies can be approximated by a 2D axisymmetric density profile. The surface brightness of many spiral galaxies can be described by an exponential profile \citep{sersic63,freeman70,courteau96}. \citet{smit17} recently report the observation of two $z\approx 6.8$ galaxies with a clear velocity gradient in \CII, which might suggest the presence of a turbulent, rotation-dominated disc. Based on imaging spectroscopy of the H$\alpha$ emission line, \citet{genzel17} confirm that the observed velocity profiles of high-z disc galaxies favour a thick exponential disc profile and \citet{pawlik11} also find a disc structure for the first galaxy in their 3D hydrodynamical simulations. We therefore assume an exponential profile for the surface density
\begin{equation}
 \Sigma (R) = \Sigma _0 \exp (-R/R_0),
\end{equation}
where $R_0$ is the scaling radius, given by \citep{mo98}:
\begin{equation}
R_0 = \frac{\lambda}{\sqrt{2}} \left( \frac{j_d}{m_d} \right) R_\mathrm{vir},
\end{equation}
with $j_d$ being the fraction of the specific angular momentum of the halo that goes into the disc, $m_d$ the fraction of the halo mass that settles into the disc, and $\lambda$ the spin parameter of the halo. For typical halos we assume $j_d \approx m_d \approx \lambda \approx 0.05$ and hence
\begin{equation}
R_0 \approx 0.035 R_\mathrm{vir}.
\label{eq:ScalingRadius}
\end{equation}
The scale height of an isothermal, self-gravitating disc is given by
\begin{equation}
H(R) = \frac{c_s ^2}{\pi G \Sigma (R)} = \frac{c_s ^2}{\pi G \Sigma _0} \exp (R/R_0),
\end{equation}
where $c_s$ is the sound speed of the gas. The derivation of this relation assumes a locally constant surface density and the real scale height might be smaller in regions where the disc surface density changes significantly with radius.
For the total density profile of the disc we require
\begin{equation}
 \Sigma (R) = \int _{- \infty} ^{\infty} \rho (R,h) \mathrm{d}h.
\end{equation}
By assuming
\begin{equation}
 \rho (R,h) \propto \cosh ^{-2} (h/H(R))
\end{equation}
we find
\begin{equation}
 \rho (R,h) = \frac{\Sigma _0}{2 H(R)} \exp (-R/R_0) \cosh ^{-2} (h/H(R)).
\label{eq:rhoexp}
\end{equation}
The gas mass inside a radius $R$ for the exponential disc is given by
\begin{equation}
M(<R) = 2 \pi \Sigma _0 R_0 \left[ R_0 - \exp (-R/R_0) (R+R_0) \right].
\end{equation}
For reasonable values of the spin parameter $\lambda$, the second term is negligible. Assuming that $m_d M_h$ is the mass of the disc within $R_\mathrm{vir}$ we can calculate the normalization of the surface density as
\begin{equation}
\Sigma _0 = \frac{m_d M_h}{\pi (\lambda R_\mathrm{vir})^2},
\end{equation}
where $M_h$ is the mass of the halo.

\subsubsection{Dark matter}
Following \citet{costa14}, we assume a spherically symmetric Hernquist profile \citep{hernquist90} of the form
\begin{equation}
\rho _{DM} = (1-f_\mathrm{gas})\frac{M_h}{2 \pi}\frac{a}{r(r+a)^3}
\end{equation}
and
\begin{equation}
M_{DM} (<r) = (1-f_\mathrm{gas}) M_h \frac{r^2}{(r+a)^2},
\end{equation}
with $f_\mathrm{gas}=\Omega_b/\Omega_m=0.17$ and the scaling radius $a=0.1R_\mathrm{vir}$. With small $r$ we denote the radius in spherical geometry in contract to capital $R$ which we use for the radius in the disc plane in cylindrical coordinates.

\subsubsection{1D reference model}
To highlight the differences of a 2D model, we also compare our new model to a spherical 1D gas distribution for which we assume that the radial gas profiles follows a Hernquist profile with
\begin{equation}
\label{eq:HernquistGas}
\rho _\mathrm{gas} = \frac{f_\mathrm{gas}M_h}{2 \pi}\frac{a}{r(r+a)^3},
\end{equation}
where $a=0.1R_\mathrm{vir}$ is the scaling radius.

\subsubsection{Fiducial galaxy model}
For our fiducial model we assume typical parameters of a $z=6$ galaxy with a halo mass of $M_\mathrm{h}=10^8\Msun$. The BH has a mass of $M_\mathrm{BH}=10^5\Msun$ and shines at an Eddington ratio of $f_\mathrm{Edd}=0.3$.
We assume that the AGN shines for a time of $t_\mathrm{on}=10$\,Myr with a constant luminosity that is given by the BH mass and the Eddington ratio. The resulting surface density normalization and disc scale height at the scale radius are $\Sigma_0 = 0.28\,\mathrm{g}\,\mathrm{cm}^{-2}$ and $H(R_0)=2.6$\,pc, respectively. In section \ref{sec:para} we vary all these model parameters independently to analyse their influence on the outflow dynamics.

We do not include a stellar population in the galaxy because its gravitational potential has no direct influence on the outflow dynamics. Galaxies at $z \geq 6$ have a stellar mass fraction of at most $10\%$, relative to their baryon content \citep{oshea15,trebitsch17}. Low-mass galaxies with a halo mass of $M_h \approx 10^8 \Msun$, such as our fiducial model, have even lower stellar mass fractions of $0.1-1\%$ \citep{behroozi13,trebitsch17}. In one test scenario we include a stellar population as a Hernquist profile with a scaling radius of $a_*=5$\,pc \citep{park16} with a stellar mass fraction of $10\%$. Even under this conservative assumption the time evolution of the AGN-driven outflow is not distinguishable from another model, in which we reduce the gas mass by 10\%. This is because the dynamics of the outflow is dominated by the inertia of the swept-up gas and not by the gravity of the external potential. Phrased differently, the inclusion of a static stellar potential does not affect the outflow and only the implicated lower gas mass has a small effect.

A smaller gas mass fraction, e.g. due to star formation at lower redshift, would decrease the surface density normalization. Also, gas in high-z galaxies is turbulent and has significant pressure support \citep{genzel08,fs09,ss12}, which would additionally increase the disc scale height. These effects would lower the ISM density in the disc plane and permit the AGN-driven wind to propagate to larger radii.

\subsubsection{Initial and final radii}
We start the integration of the outward driven shock at an initial radius $R_\mathrm{min}$. The physical interpretation of this radius is the innermost distance where the disc wind starts to sweep up the interstellar medium (ISM). We use the self-gravity radius of the accretion disc as a proxy for $R_\mathrm{min}$ and we express it in units of the Schwarzschild radius:
\begin{equation}
R_\mathrm{SS} = \frac{2GM_\mathrm{BH}}{c^2} \approx 3\times 10^5 \frac{M_\mathrm{BH}}{\Msun}\,\mathrm{cm}.
\end{equation}
For standard values of the disc viscosity and Eddington ratio the outer radius of the disc (i.e. self-gravity radius) and hence the inner radius for our integration is given by \citep{ss73,goodman04,king08,dotti10}
\begin{equation}
R_\mathrm{min} \approx 10^{5} R_\mathrm{SS},
\end{equation}
which yields consistent results for all the tested parameter combinations and for BH masses, spanning many orders of magnitude.

We integrate the equations out to arbitrary radii, but the physically interesting regime is within the virial radius. Hence, we follow the outflow in each direction until it either becomes subsonic, or until it crosses the virial radius of the halo. Since the escape velocity at the virial radius is roughly equal to the sound speed, we assume that the wind escapes the gravitational potential of the halo if it reaches the virial radius and is still supersonic.

\subsection{Quantifying the outflow}
We define a set of parameters that quantify the efficiency, dynamics, and nature of the outflow. 

We introduce a radius, $R_\mathrm{max}$, out to which the outflow has to push the gas at least to stop further gas accretion on to the central BH. Due to the higher gas density in the disc plane, it is a sufficient criterion for the outflow to reach $R_\mathrm{max}$ in the disc plane. A sufficient criterion to prevent further gas accretion on to the central BH is $R_\mathrm{max} = R_\mathrm{vir}$ and a necessary criterion is $R_\mathrm{max}>R_\mathrm{soi}$, where $R_\mathrm{soi}$ is the sphere of influence, within which the dynamics is dominated by the BH.

Once pushed out to a certain radius, the question is if this gas can reach the BH in a sufficiently short amount of time to start refuelling the AGN. We hence relate $R_\mathrm{max}$ to the recovery time $t_\mathrm{ff}$, which we define as the minimum time the gas needs to fall back towards the central BH, after being pushed out to the radius $R_\mathrm{max}$.
Expressing the free-fall time via the sound crossing time yields for $R_\mathrm{max}$ in the disc plane
\begin{equation}
t_\mathrm{ff} = 5\,\mathrm{Myr} \left( \frac{R_\mathrm{max}}{R_0} \right) \left( \frac{1+z}{7} \right)^{-3/2}.
\label{eq:Rff}
\end{equation}
There is no unique criterion on the absolute value of $R_\mathrm{max}$ or $t_\mathrm{ff}$ to shut off further gas accretion. We therefore present and discuss both relative to the scale radius of the disc and to the lifetime of the AGN. If $t_\mathrm{ff}$ is greater than the AGN lifetime, we can assume that the outflow is efficient enough to significantly suppress further mass accretion.

The ejected mass $M_\mathrm{eject}$ is the gas mass that is ejected from the gravitational potential of the halo, i.e. that passes the virial radius with a velocity larger than the escape velocity. The ejection angle $\theta_\mathrm{eject}$ is the last angle (measured from the disc normal) for which gas can escape the halo. It defines the opening angle of a cone, which is cleared off gas by the AGN-driven wind. Gas outside this cone is pushed to $R_\mathrm{max}(\theta)$, but falls back towards the galactic centre over a time $\gtrsim t_\mathrm{ff}$. The free-fall time in the disc plane is defined as the time which gas that has been pushed to a certain radius needs at least to fall back towards the central BH, once the shock becomes subsonic (see Eq. \ref{eq:Rff}). This is an approximate quantification for how long an AGN can shut off its own gas supply.

We also compare the mass accretion rate of the central BH:
\begin{equation}
\dot{M}_\mathrm{acc} = \frac{L}{\eta c^2},
\label{eq:Macc}
\end{equation}
to the mass outflow rate:
\begin{equation}
\dot{M}_\mathrm{out} = \sum _i M_\mathrm{shell,i} \frac{v_\mathrm{shell,i}}{R_\mathrm{shell,i}},
\label{eq:Mout}
\end{equation}
where the mass and velocity of the shell resolution elements are summed over all angles $i$. The mass outflow rate generally varies with time, but to make it more accessible for a direct comparison to the mass accretion rate, we define the mass outflow rate at the sphere of influence, which is one unique number. In theory, the shell crosses the sphere of influence at different times for different angles and we need to sum over shell elements at different ``crossing-times''. However, in practice, the shell crosses the sphere of influence for all angles in the same time-step, because the density profile of the exponential disc is close to spherical in the centre and there is no preferred direction. Comparing these two mass rates quantifies the efficiency of the AGN to convert infalling, accreted mass into a mass outflow.

The ratio $R_\mathrm{perp}/R_\mathrm{disc}$ quantifies the asymmetry of the outflow. At any time, $R_\mathrm{disc}$ is the position of the shock front in the disc plane and $R_\mathrm{perp}$ perpendicular to it. If this ratio is equal to one, the wind propagates almost spherically symmetric. For values above unity the outflow develops peanut-shaped lobes. This ratio illustrates the importance and improvement of a 2D treatment with respect to 1D wind solutions.

We define the momentum boost, which is often referred to as ``mechanical advantage'', as the ratio of the total momentum of the AGN-driven wind divided by the total momentum input by photons:
\begin{equation}
\frac{\sum _i |\textbf{p}_i|}{tL/c}.
\end{equation}
The vectorial sum would be equal to zero, but summing the absolute values of the momentum allows to quantify the efficiency of the AGN to accelerate the gas. In the momentum-driven regime we expect
\begin{equation}
\sum _i |\textbf{p}_i| \leq tL/c,
\end{equation}
but the adiabatic expansion in the energy-driven case can yield mechanical advantages above unity. Observations find momentum boost in the range of $\sim$ 2--30 \citep{rupke11,sturm11} and 3D simulations yield momentum boosts of 1--30 \citep{cicone14,bieri17}.

\subsection{Shock acceleration}
There are two systematically different types of outflows: if the shocked wind cools efficiently, the outflow is accelerated only by the momentum transfer and the outflow is called \textit{momentum-driven}. A momentum-driven outflow develops a thin shock layer. For less efficient cooling, the shock-heated gas remains at temperatures of $>10^7$\,K and the adiabatic expansion of the hot shocked wind accelerates the shock into the ISM. In this case, a thick shock layer develops and the outflow is labelled \textit{energy-driven}, because the internal energy of the adiabatically expanding shock-heated wind drives the outflow.

In a disc-like geometry we expect the shock to propagate faster into the direction of lower density. This implies that a self-consistent 2D treatment of the wind propagation is important to capture all the relevant physics of the shock dynamics.

We can assume that the shock front is locally plane-parallel and in the rest frame of the shock front, the Rankine--Hugoniot jump conditions for a plane parallel shock are as follows:
\begin{align}
\rho _1 u_1 &= \rho _2 u_2 \\
\rho _1 u_1 ^2 +P_1 &= \rho _2 u_2 ^2 +P_2 \\
\frac{1}{2} \rho _1 u_1 ^2 + \epsilon _1 +\frac{P_1}{\rho _1} &= \frac{1}{2} \rho _2 u_2 ^2 + \epsilon _2 +\frac{P_2}{\rho _2},
\end{align}
where the index 1 refers to the pre-shock and the index 2 to the post-shock conditions of the density $\rho$, velocity $u$, pressure $P$, and the specific internal energy $\epsilon$.
For a strong shock with Mach number
\begin{equation}
\mathcal{M}_1 = \left( \frac{\rho _1 u_1 ^2}{\gamma P_1} \right) ^{1/2} \gg 1,
\end{equation}
we find
\begin{align}
\label{eq:rhopost}
\frac{\rho_2}{\rho _1} &\approx \frac{\gamma + 1}{\gamma - 1} = 4 \\
P_2 &\approx \frac{2}{\gamma +1} \rho _1 u_1 ^2 = \frac{4}{3} \rho _1 u_1 ^2 \\
\label{eq:Tpost}
T_2 &\approx \frac{2(\gamma -1)}{(\gamma +1)^2} \frac{m}{k_\mathrm{B}} u_1 ^2 = \frac{3}{16} \frac{m}{k_\mathrm{B}} u_1 ^2\\
\mathcal{M}_2 &\approx \left( \frac{\gamma -1}{2 \gamma} \right) ^{1/2} \approx 0.45,
\end{align}
where the last equalities are valid for an adiabatic coefficient of $\gamma=5/3$ \citep{shull87}. Hence, a shock converts supersonic gas into denser, slower moving, higher pressure, subsonic gas.

The density of the shocked wind depends on the available cooling channels. For a radiative (momentum-driven) strong shock, the post-shock density is given by $\rho _2 = 4 \rho _1$ (Eq. \ref{eq:rhopost}). In the energy-driven case the density of the shocked wind depends on the thickness of the shock. We generally refer to the shock position as the contact discontinuity between the shocked wind and the shocked ISM, and the thickness of the shocked wind is given by
\begin{equation}
\label{eq:dR}
\Delta r = r - r_\mathrm{sw},
\end{equation}
where $R$ and $R_\mathrm{sw}$ are the position of the contact discontinuity and of the wind shock, respectively \citep[see illustrations of outflow dynamics in][]{zubovas12,FG12,king15,dashyan17}. The radius where the wind is shocked for an energy-driven outflow is given by \citet{FG12}:
\begin{equation}
\label{eq:Rsw}
r_\mathrm{sw} \approx r \sqrt{\frac{v}{v_\mathrm{in}}},
\end{equation}
and the thickness of the shocked wind is $\Delta r = r (1- (v/v_\mathrm{in})^{1/2})$. 
We do not use this thickness of the shock explicitly for the integration of the equations of motion, but we use it implicitly to calculate the rate of radiative cooling (Sec. \ref{sec:radcool}).


\subsubsection{Equation of motion: momentum-driven}
The shock is accelerated by the momentum input of the AGN, counteracted by gravity \citep{king10,ishibashi14,ishibashi15,king15}:
\begin{align}
&\frac{\mathrm{d}}{\mathrm{d}t} \left[ M_\mathrm{shell}(r) \dot{r} \right] = \frac{f_\mathrm{edd} L_\mathrm{edd}}{c} \nonumber \\
&-G \frac{M_\mathrm{shell}(r)(M_\mathrm{DM}(<r)+M_\mathrm{BH})}{r^2} - F_\mathrm{selfgrav},
\label{eq:expansion}
\end{align}
where $f_\mathrm{edd}$ is the Eddington fraction, $M_\mathrm{shell}$ is the mass of the swept-up ISM, and
\begin{equation}
L_\mathrm{edd} = \frac{4 \pi c G M_\mathrm{BH} m_p}{\sigma _T}
\end{equation}
is the Eddington luminosity. The term $F_\mathrm{selfgrav}$ is not explicitly mentioned in most studies \citep{king10,ishibashi14,ishibashi15,king15}, but it accounts for the swept-up gas in the shell and the gravity contribution from the gas outside the shell. The gravitational force of the ambient gas is negligible, but self-gravity of the shell has to be included.


The term $L/c$ describes the total momentum per unit time that can be transferred to the gas via single scatter events. For Thomson scattering we assume that the directions of the photons after one scattering are random and that the momentum transfer from secondary scattering events cancels out.
However, there is a probability $0 < p \leq1$ for the photons to escape freely without any interaction with the ISM: 
\begin{equation}
\ln (p)= - \frac{\sum _i \sigma _{t,i}}{d \Omega r^2} = - \frac{M_\mathrm{shell}}{m_p} \frac{\sigma _t}{d \Omega r^2} = - \kappa \Sigma _\mathrm{shell} = - \tau _\mathrm{shell},
\end{equation}
with the opacity $\kappa = \sigma _t / m_p = 0.4 \mathrm{cm}^2\,\mathrm{g}^{-1}$.
The fraction of the total force input $L/c$ that actually couples to the gas is hence
\begin{equation}
\frac{L}{c} \left( 1- e^{- \tau_\mathrm{shell}} \right).
\end{equation}
We do not explicitly account for this effect and show in Section \ref{sec:fiducial} that it is negligibly small and irrelevant in most of the scenarios considered. However, see \citet{ishibashi15} for a more detailed discussion of this effect and for the influence of multiscattering of IR photons on dust grains.

\subsubsection{Equation of motion: energy-driven}
In the energy conserving case, the shocked ISM is not accelerated by the direct momentum input, but by the adiabatic expansion of the shocked wind. The momentum equation then reads
\begin{equation}
\frac{\mathrm{d}}{\mathrm{d}t}[M_\mathrm{shell}(r)\dot{r}]+F_\mathrm{grav}=4 \pi r^2 P,
\end{equation}
where $P$ is the pressure in the shocked wind, which can be calculated via energy conservation
\begin{equation}
(\gamma -1) \frac{\mathrm{d}}{\mathrm{d}t} (PV) = \frac{\eta}{2} f_\mathrm{Edd} L_\mathrm{Edd} - P \frac{\mathrm{d}}{\mathrm{d}t}V - F_\mathrm{grav} \dot{r},
\label{eq:energy}
\end{equation}
where $V$ is the volume and $\eta$ sets the mechanical luminosity, which can be expressed as $\eta = v_\mathrm{in}/c$ (see section \ref{sec:vin}). This energy conservation assumes that the thermal energy of the shocked wind is much higher than its kinetic energy, which is generally the case \citep{FG12}.


\subsection{Cooling and Heating}
\label{sec:cool}

In this section we discuss the most important heating and cooling mechanisms of the gas. We demonstrate that inverse Compton cooling is the only relevant mechanism on the scales of interest \citep{ciotti97} and discuss why radiative cooling of the wind shock can be neglected. 

\subsubsection{Compton Cooling}
The Compton cooling time of gas is given by
\begin{equation}
t_\mathrm{compton} = \frac{3 m_e c}{8 \pi \sigma _T U_\mathrm{rad}} \frac{m_e c^2}{E},
\end{equation}
with the radiation energy density
\begin{equation}
U_\mathrm{rad} = \frac{f_\mathrm{Edd} L_\mathrm{Edd}}{4 \pi R^2 c}
\end{equation}
and the internal energy in the shocked gas
\begin{equation}
E= \frac{9 m_p v_\mathrm{in}^2}{16},
\end{equation}
where $v_\mathrm{in}$ is the initial velocity of the disc wind, before it shocks with the ISM (see Sec. \ref{sec:vin}) and $m_p$ and $m_e$ are the masses of the proton and electron, respectively. The Compton cooling time is hence given by \citep[see also][]{FG12}
\begin{equation}
t_\mathrm{compton} = \frac{2}{3 \pi f_\mathrm{edd}} \frac{c}{G M_\mathrm{BH}} \left( \frac{m_e}{m_p} \right) ^2 \left( \frac{v_\mathrm{in}}{c} \right) ^{-2} r^2.
\label{eq:compton2}
\end{equation}

\subsubsection{Two temperature medium}
\label{sec:twot}
Inverse Compton cooling acts only on the electrons. \citet{FG12} analyse the cooling properties of a medium with different temperatures of the electrons and ions. They determine the characteristic time-scale to achieve thermal equilibrium ($T_e = T_p$) to be
\begin{equation}
t_{ei} = \frac{3 m_e m_p}{8 (2 \pi)^{1/2} n_p e^4 \ln \Lambda} \left( \frac{k_B T_e}{m_e} \frac{k_B T_p}{m_p} \right)^{3/2},
\end{equation}
where $e$ is the elementary charge and $\Lambda \approx 40$. Assuming $T_p = 10T_e = 10^{10}$\,K and a proton density of $n_p = 1 \mathrm{cm}^{-3}$ as typical post-shock conditions, this time-scale is of the order $t_{ei} \approx 1$\,Myr. \citet{FG12} show that this time-scale is longer than the Compton cooling time and might delay the cooling by up to two orders of magnitude. We do not include this effect of a two-temperature medium in our calculations and discuss its effect in Sec. \ref{sec:cav}.


\begin{figure}
\centering
\includegraphics[width=0.47\textwidth]{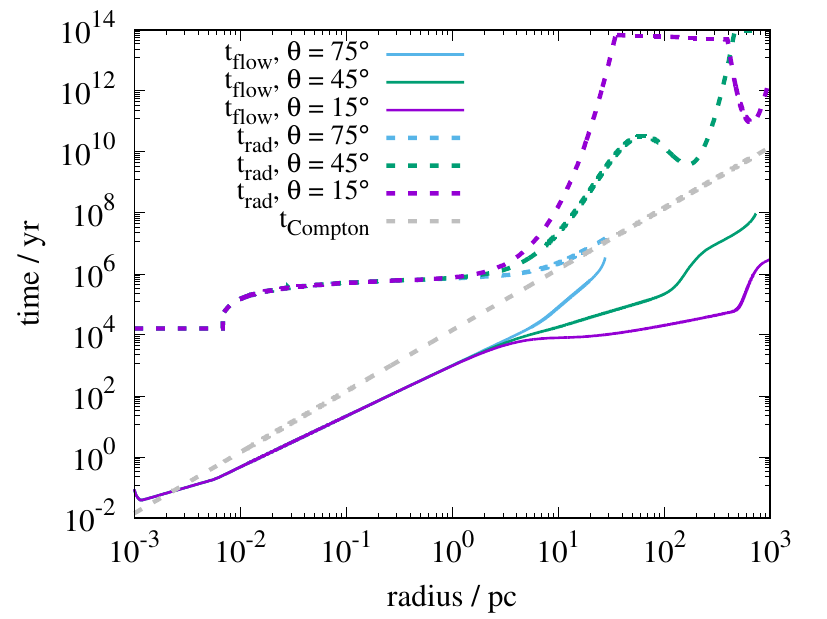}
\caption{Radial profiles of characteristic time-scales for our fiducial model, where the flow time is defined as $t_\mathrm{flow}=R/v$. The different colours indicate different angles with respect to the disc normal, $\theta$, and the Compton cooling time in this radial range is independent of the angle (grey).}
\label{fig:rad}
\end{figure}

\subsubsection{Radiative Cooling}
\label{sec:radcool}
We only consider inverse Compton cooling in our calculation of the cooling time. To quantify the possible contribution of radiative cooling, we use the cooling function by \citet{sutherlandDopita} in the functional form by \citet{tozzi01}.
The post-shock temperature is given by Eq. \ref{eq:Tpost} and we calculate the gas density based on Eq. \ref{eq:dR} or Eq. \ref{eq:rhopost}, depending on whether the shock is energy- or momentum-driven, respectively. The radiative cooling time becomes
\begin{equation}
t_\mathrm{rad} = \frac{9 v_\mathrm{in}^2 \mu m_\mathrm{p}^2}{32 \Lambda \rho _\mathrm{shock}},
\end{equation}
where $\rho_\mathrm{shock}$ is the corresponding post-shock density and $\Lambda$ the cooling function. The radial profile of the radiative cooling time and a comparison to other characteristic time-scales can be seen in Fig.\ref{fig:rad}.
Within $\sim 0.01$\,pc the velocity of the swept-up material is close to the initial wind velocity and therefore the shock would be infinitely thin (Eq. \ref{eq:Rsw}). This implies a very high density and an artificially short cooling time. In this case we set the post-shock density to $\rho_\mathrm{shock}=4\rho$ to avoid singularities. On all scales of interest, the radiative cooling time is always longer than the Compton cooling time and can be neglected.

\begin{figure}
\centering
\includegraphics[width=0.47\textwidth]{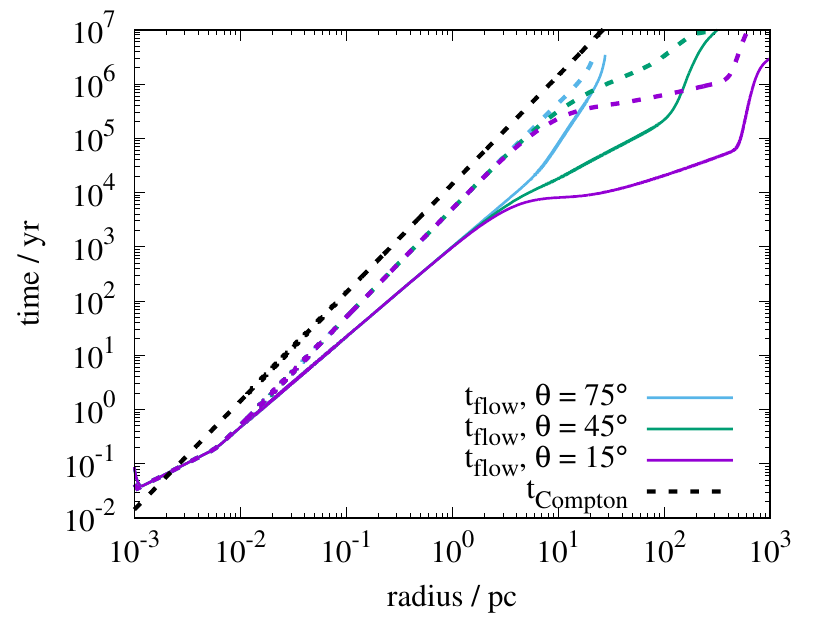}
\caption{Characteristics times as a function of radius for different angles with respect to the disc normal for our fiducial model. The black dashed line shows the Compton cooling time (Eq. \ref{eq:compton2}) and the solid lines show the flow time, for the self-consistent driving, i.e. energy-driven when $t_\mathrm{flow}<t_\mathrm{Compton}$. The dashed lines illustrate the dynamics of an outflow that is always momentum-driven. The transition from momentum- to energy-driven occurs at $\sim 2 \times 10^{-3}$\,pc and the momentum-driven solution is less efficient in driving an outflow.}
\label{fig:self}
\end{figure}

\subsection{Mechanical luminosity and initial wind velocity}
\label{sec:vin}
The post-shock temperature is set by the pre-shock wind velocity $v_\mathrm{in}$ (Eq. \ref{eq:Tpost}). By assuming that the initial wind mass rate $\dot{M}_\mathrm{in}$\,is equal to the accretion rate on to the BH $\dot{M}_\mathrm{BH}$, with a radiative efficiency of $\eta \approx 0.1$ for the AGN, we obtain
\begin{equation}
v_\mathrm{in} = \eta c \approx 30000\,\mathrm{km}\,\mathrm{s}^{-1}.
\end{equation}
The assumption $\dot{M}_\mathrm{in}=\dot{M}_\mathrm{BH}$ might not be accurate, but we use it as a working hypothesis, following e.g. \citet{FG12,costa14,king15} and we explore the dependence on $v_\mathrm{in}$ in Sec. \ref{sec:vinpara}. The assumption of $v_\mathrm{in} = 0.1c$ is supported by observations of the Doppler-shifted Fe $K$ band absorption lines, which are observed in many local AGNs \citep{pounds03,tombesi10,tombesi16}.
\citet{ostriker10} provide a more detailed and complete discussion of the momentum driving and mechanical luminosity.

This assumption enters in the calculation of the cooling time and for the mechanical input of the AGN. If we assume that the initial mass outflow rate is only a fraction of the accretion rate:
\begin{equation}
\dot{M}_\mathrm{in}=f_\mathrm{acc}\dot{M}_\mathrm{BH},
\end{equation}
we would find $v_\mathrm{in} = \eta f_\mathrm{acc} c$ for the initial velocity and the energy equation would change to
\begin{equation}
\frac{\eta}{2} L_\mathrm{AGN} \rightarrow \frac{\eta}{2} f_\mathrm{acc} L_\mathrm{AGN}  = \frac{\eta}{2} f_\mathrm{acc} f_\mathrm{Edd} L_\mathrm{Edd}.
\end{equation}
Hence, in the parameter study of the outflow dynamics, $f_\mathrm{acc}$ and $f_\mathrm{Edd}$ enter the equation of motion as a product, which allows us to keep one parameter fixed ($f_\mathrm{acc}=1$) and vary only the Eddington ratio.

We assume that this initial line-driven disc wind is isotropic. This is supported by observations of ultrafast outflows in a large fraction of observed AGNs. Unless we see all AGNs from a very particular angle, these AGN-driven winds have large opening angles with an almost spherical driving mechanism \citep{king15}.

\subsection{Momentum- to energy-driven transition}
The main problem arises since the nature of the outflow depends on the dynamics of the shock and the shock dynamics depends on the nature of the outflow. We self-consistently check at any time if the outflow is energy- or momentum-driven (see Fig.~\ref{fig:self}).

We expect the transition to happen at different radii for different angles with respect to the disc plane. This implies that there are phases where momentum- and energy-driving occur at the same time in a given galaxy. For the energy input by the AGN in driving the adiabatic expansion, we correct it by the solid angle $\mathrm{d}\Omega$ of the energy-driven outflow, because for $4\pi - \mathrm{d}\Omega$ we expect the energy input to be radiated away in the momentum-driven shock front. However, the remaining question is, which volume represents the correct basis for the energy equation of the adiabatic expansion: the entire volume enclosed by the shock front or only the volume enclosed by the energy-driven shock front? A Gedankenexperiment can help to settle this question: if we only had to consider the volume of the cone enclosed by the energy-driven shock (perpendicular to the disc plane, because we expect higher velocities in this direction), then what happens at the boundary of this cone? The adiabatically expanding volume does not know about whether it is enclosed by an energy- or momentum-driven shock front, but rather wants to expand in all directions. For the same reason, we have to adopt our treatment of the adiabatically expanding volume once the shock becomes subsonic: the gas still expands and can do so into the ISM with $\sim c_s$, even if the shock front is dissolved by turbulence. For directions into which the shock becomes subsonic, we hence follow the propagation of the swept-up mass and the adiabatically expanding volume separately. We have tested the influence of these methods and find that the difference is small.

\subsection{Validation of the model: comparison to 1D solution}
\begin{figure}
\centering
\includegraphics[width=0.47\textwidth]{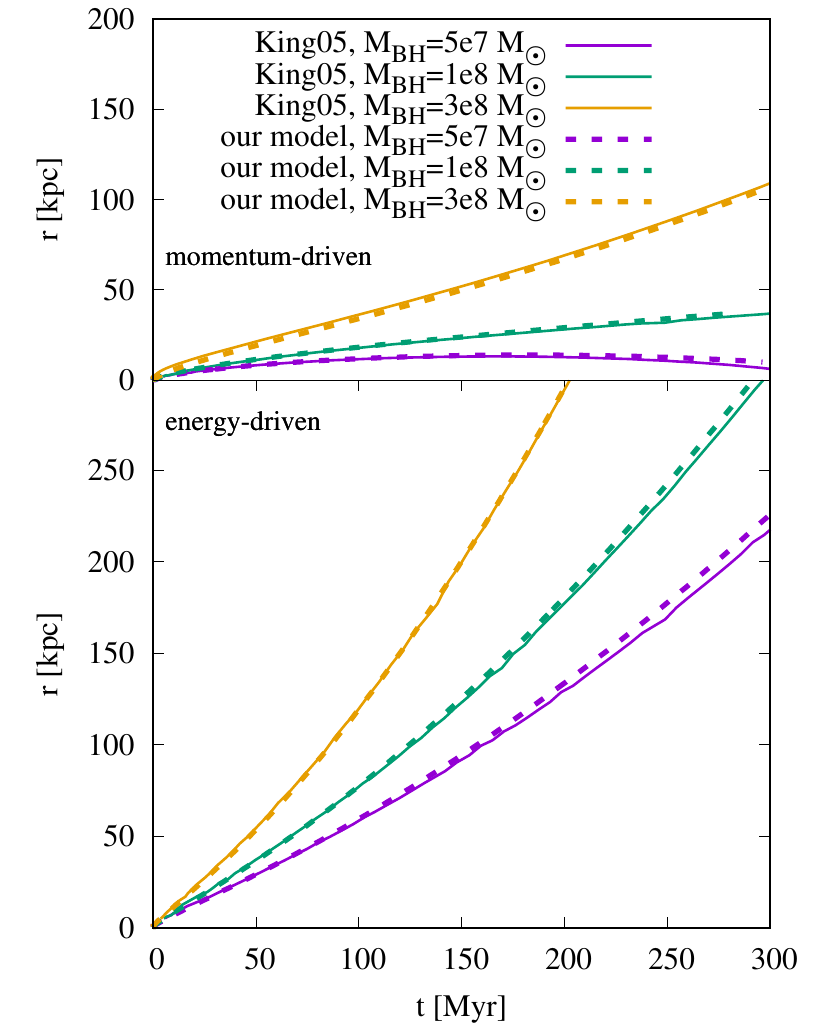}
\caption{Comparison of our model with the results by \citet{king05}, which \citet{costa14} can also reproduce with their 3D simulation, for different BH masses. Here, we assume the same spherical density distribution as \citet{costa14}, but integrate the equations of motion with our 2D model. Our results agree well with the analytical solution by \citet{king05} and with the 3D simulations by \citet{costa14}.}
\label{fig:comparison}
\end{figure}

To test the robustness and convergence of the method, we first compare the results of our 2D analytical model to the results of \citet{costa14} and \citet{king05}. We assume $M_h = 10^{12} \Msun$ for the halo mass, a Hernquist profile for the dark matter and the gas, where the gas mass is $M_\mathrm{gas}=f_\mathrm{gas}M_h$ with $f_\mathrm{gas} = 0.17$ and the scaling radius $a=28$\,kpc. The virial radius is $R_\mathrm{vir}=163$\,kpc, the AGN shines at the Eddington luminosity and for the BH we assume three different masses of $M_\mathrm{BH}=5\times10^7,10^8,3\times 10^8 \Msun$. We assume that there is no gas within the sphere of influence of the BH, which is $\sim 500$\,pc for these configurations. The results can be seen in Fig.~\ref{fig:comparison}.

\begin{figure}
\centering
\includegraphics[width=0.47\textwidth]{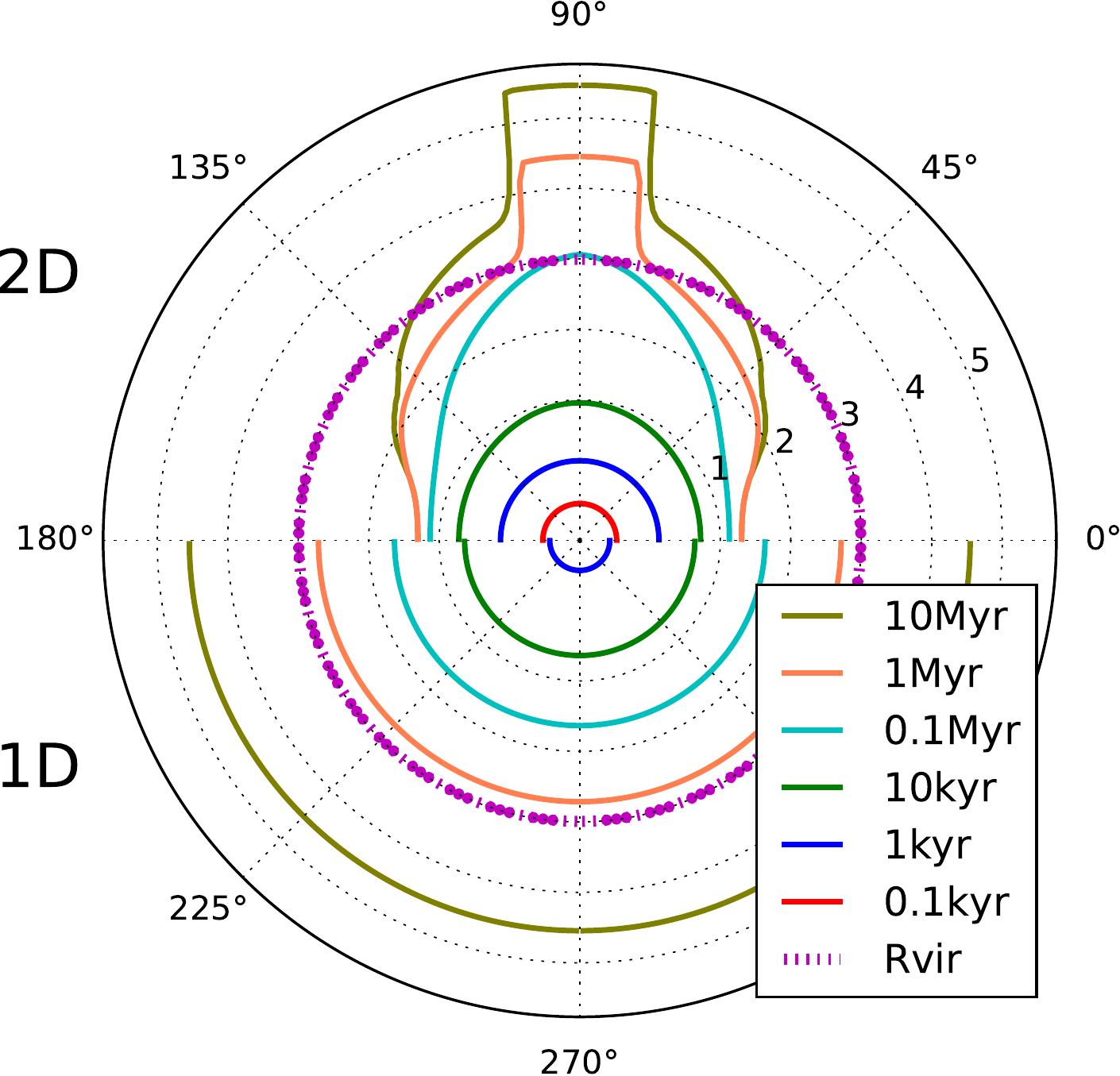}
\caption{Time evolution of the position of the shock front in polar coordinates. Top: 2D solution with the galactic disc plane at $0^{\circ}$. Bottom: 1D spherical solution. The concentric grey rings indicate the radius in $\log(r/\mathrm{pc})$ from 10\,pc to 100\,kpc and the magenta circle represents the virial radius at $980$\,pc. In the disc plane, the shock cannot propagate beyond a maximal radius of $\sim 20$\,pc and only gas at $\gtrsim 47^{\circ}$ escapes the virial radius. The 1D wind is slower at $\lesssim 30$\,kyr, but ejects more gas out of the virial radius at later times.}
\label{fig:shells}
\end{figure}

Our 2D analytical model is able to reproduce the outflow dynamics of the 1D model by \citet{king05} and of the 3D simulation by \citet{costa14} in a spherical gas profile. We then adopt the same functional forms for the gas distribution (Hernquist profile) and apply it to our fiducial model with $M_h=3\times 10^7\Msun$ and $M_\mathrm{BH}=10^5\Msun$. The direct comparison of the outflow dynamics in 1D and 2D is given in Fig.~\ref{fig:shells}.
For the same gas mass and AGN luminosity, the outflow in the 2D model can only eject mass within a cone with an opening angle of $\sim 45^{\circ}$, whereas the same outflow in the 1D model ejects all the gas out of the virial radius. The main reason for this lower efficiency in the 2D scenario is that the pressure in the 1D model cannot escape, but builds up over time. In the 2D scenario, however, the pressure decreases with time due to the fast expansion of the energy-driven wind perpendicular to the disc.

\section{Results}

\subsection{Standard case (fiducial parameters)}
\label{sec:fiducial}
\begin{figure}
\centering
\includegraphics[width=0.47\textwidth]{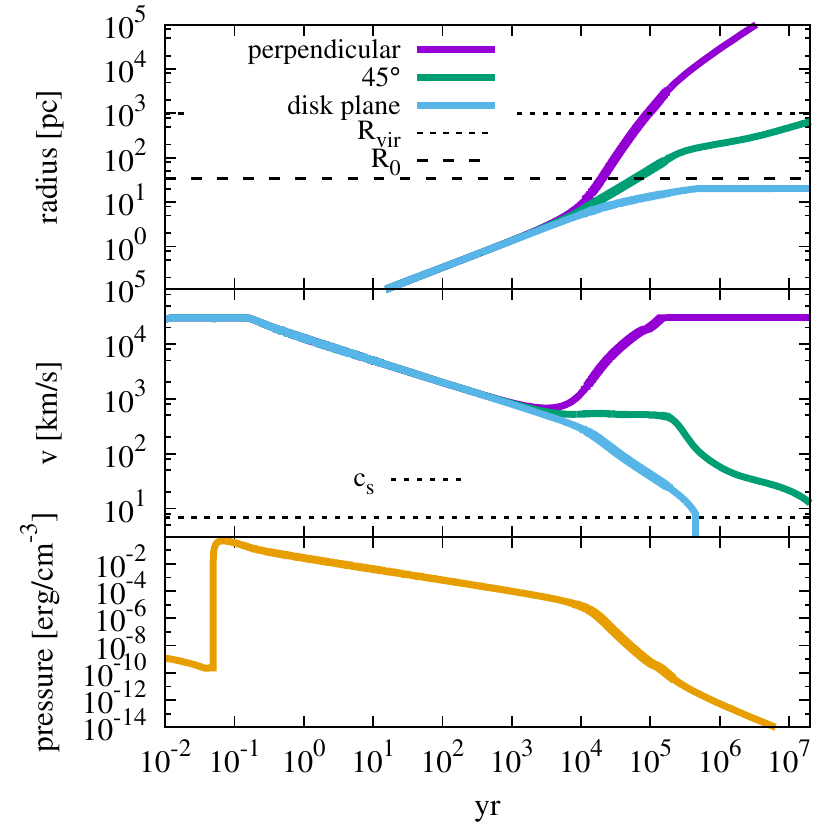}
\caption{Time evolution of different quantities for our fiducial galaxy model. Top panel: position of the shock for different angles with respect to the disc. The virial and disc scaling radius are indicated by the black dashed lines. Middle panel: velocity for different angles. The wind becomes subsonic in the disc plane at around $0.5$\,Myr. Lower panel: the pressure of the shocked wind is isotropic and it rises rapidly once the gas can no longer cool efficiently.}
\label{fig:trvp}
\end{figure}

We show the time evolution of the main dynamical quantities for our fiducial model in Fig.~\ref{fig:trvp}.
Within $\sim 0.01$\,pc the swept-up mass and hence the gas inertia is negligibly small and the dynamics is dominated by the impinging disc wind, which has an initial velocity of $v_\mathrm{in}=0.1c$ in our fiducial model. Our model is insensitive to the choice of the initial radius of integration $R_\mathrm{min}$, as long as it is within this disc wind-dominated region. When the cooling time becomes longer than the flow time, the pressure rises by about ten orders of magnitude and the outflow becomes energy-driven. The outflow slows down due to the inertia of the swept-up ISM, which was at rest first. Within $\sim 10$\,pc the density profile is almost spherically symmetric, but we can see a clear difference for the outflow beyond this radius for different angles with respect to the disc plane. While the shock front in the disc plane cannot escape the virial radius, the gas perpendicular to the disc plane gets ejected out of the galaxy.

\begin{figure}
\centering
\includegraphics[width=0.47\textwidth]{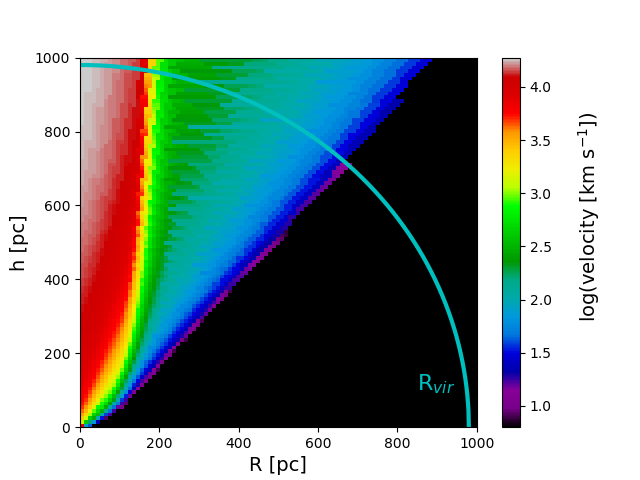}
\newline
\includegraphics[width=0.47\textwidth]{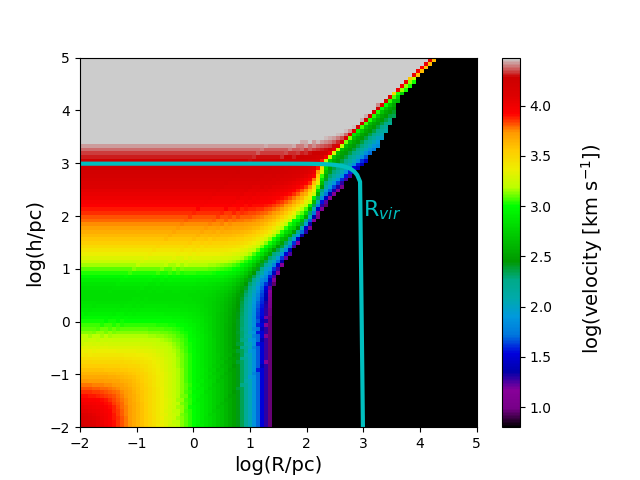}
\caption{Velocity maps for our fiducial galaxy model with the galactic disc along the x-axis and the virial radius highlighted in cyan. The linear scaling in the top panel illustrates nicely the overall geometry and the logarithmic scaling in the lower panel highlights different zones of interest: at small radii, the wind starts with a high velocity, but is decelerated by the increasing mass of the swept-up gas. Whereas the velocity is a decreasing function with radius in the disc plane, the shock accelerates in the direction perpendicular to the disc beyond $\sim 5$\,pc due to the strongly decreasing gas density in this direction.}
\label{fig:velMap}
\end{figure}
\begin{figure}
\centering
\includegraphics[width=0.47\textwidth]{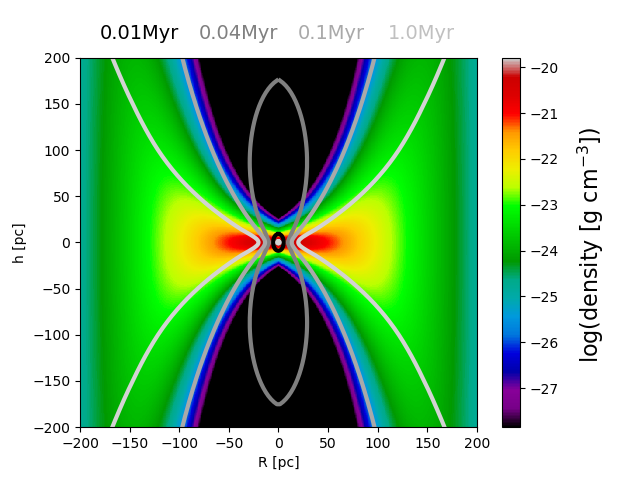}
\caption{Time evolution of the shock front with the initial density distribution in the background. The shock front starts spherically symmetric but it quickly develops an elongated shape in the direction of least resistance, i.e. perpendicular to the disc plane. The virial radius for the galaxy is $\sim 1$\,kpc.}
\label{fig:densMap}
\end{figure}

We further illustrate the outflow dynamics in Fig.~\ref{fig:velMap}.
Depending on the angle, the outflow has typical velocities of $100$--$10000\,\mathrm{km}\,\mathrm{s}^{-1}$ in our fiducial model. The bottom panel nicely shows how the outflow is re-accelerated in the perpendicular direction beyond $\sim 5$\,pc due to the exponentially decreasing ISM density. We show the time evolution of the shock structure in Fig.~\ref{fig:densMap}.
The outflow starts spherically symmetric, but develops an elongated shape, perpendicular to the disc plane after $>10$\,kyr. After $\sim 1$\,Myr, the outflow has stopped in the disc plane in our fiducial model and keeps propagating along the poles.

\begin{figure}
\centering
\includegraphics[width=0.47\textwidth]{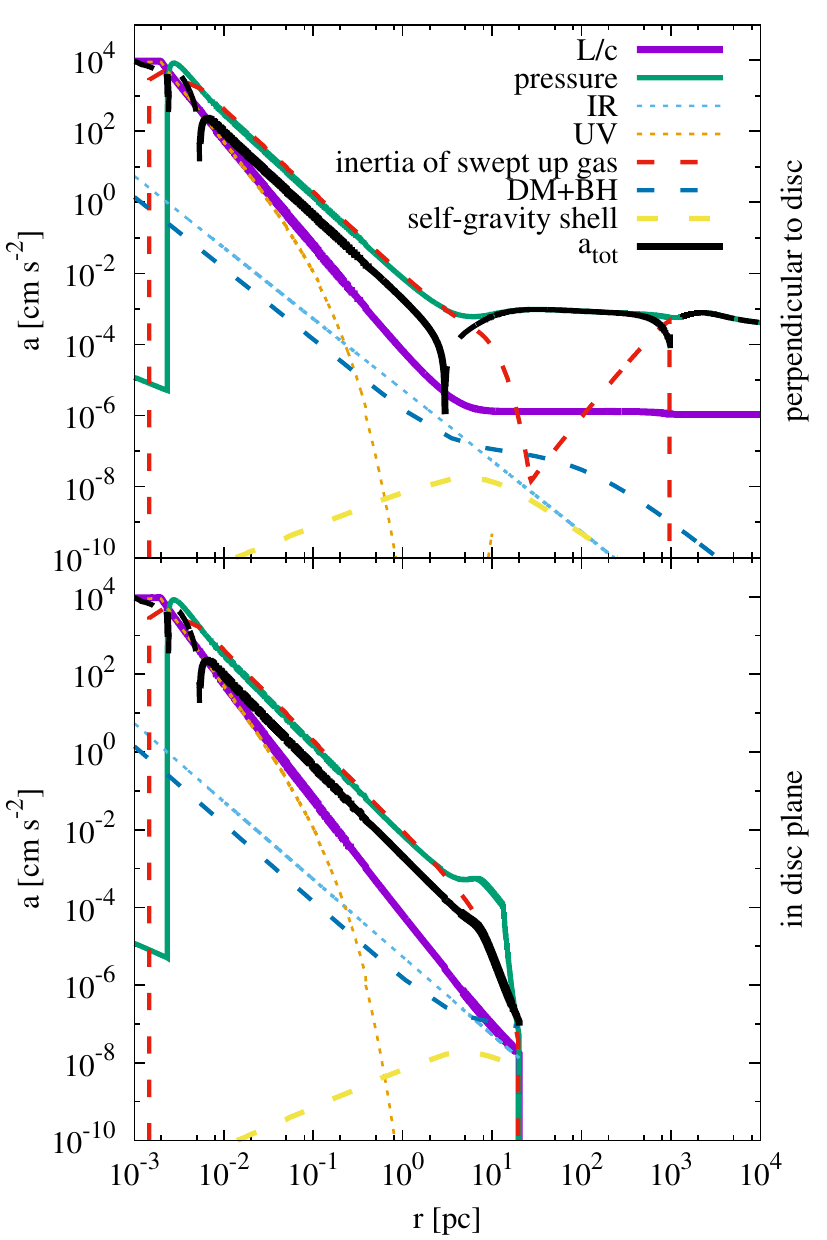}
\caption{Accelerations as a function of radius for the different contributions perpendicular to the disc (top) and in the disc plane (bottom) for our fiducial galaxy model. Solid lines are outward and dashed lines are inward contributions. Dotted lines represent accelerations that were calculated a posteriori and are not included self-consistently in the dynamics. The black line represents the total acceleration. The dynamics is dominated by the equilibrium between acceleration due to the AGN and the inertia of the swept-up ISM.}
\label{fig:acc}
\end{figure}

To analyse the influence of the individual driving and restoring forces, we plot the different contributions to the effective acceleration on a shell element in Fig.~\ref{fig:acc}.
The main forces are the acceleration by the AGN (either $L/c$ or the pressure of the adiabatically expanding shell) and the inertia of the swept-up gas, which has to be accelerated. The gravity from the BH and DM, as well as the self-gravity of the swept-up gas are negligibly small on all scales of interest. The total acceleration changes sign several times: within $\lesssim~0.01$\,pc the wind accelerates the shell until the mass of the newly swept-up material becomes significant. Within $\lesssim 5$\,pc, the wind decelerates due to the inertia of the swept-up gas. Perpendicular to the disc, the wind accelerates again at $\gtrsim~5$\,pc due to the decreasing gas density and hence smaller gas inertia. In the disc plane, the velocity is too small and the shock becomes subsonic. The deceleration and subsequent acceleration perpendicular to the disc plane around $\sim 5$\,pc can also be seen in Fig.~\ref{fig:velMap}.

The contribution of IR and UV account for absorption in the UV and re-emission in the IR, following \citet{ishibashi15} with the IR and UV opacities $\kappa _\mathrm{IR}=5\,\mathrm{cm}^2\,\mathrm{g}^{-1}$ and $\kappa _\mathrm{UV}=10^3\mathrm{cm}^2\,\mathrm{g}^{-1}$. Even for solar metallicity, these effects are subdominant compared to other contributions, but most importantly they are only relevant in the momentum-driven regime, where the momentum of the photons couples directly to the gas. See also \citet{bieri17} for a more detailed discussion of the contribution of different photon groups to AGN feedback.

\subsection{Parameter Study}
\label{sec:para}
The advantage of our analytical model is that we can easily explore large sets of parameters to study the dependence of the AGN-driven wind on e.g. the BH mass, redshift, or the AGN lifetime. In this section, we present different parameter studies and analyse their effect on the dynamics of the outflow. Some of the tested parameter combinations represent extreme scenarios and might be very rare in nature or not appear at all. We will also comment on the likelihood of the presented parameter choices.

\subsubsection{AGN lifetime}
\begin{figure}
\centering
\includegraphics[width=0.47\textwidth]{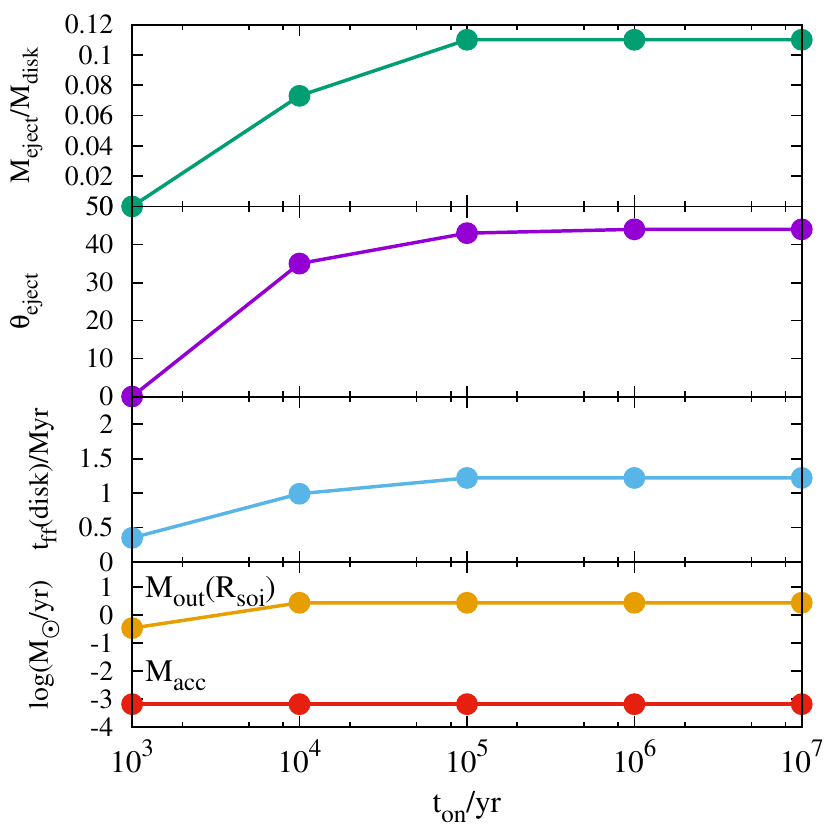}
\caption{Dependence of different quantities on the lifetime of the AGN. The ability of the outflow to remove gas from the galaxy increases with $t_\mathrm{on}$ up to maximum values above $t_\mathrm{on}\geq 0.1$\,Myr. The mass accretion and mass outflows rate are defined in Eq. \ref{eq:Macc} and \ref{eq:Mout}.}
\label{fig:ton}
\end{figure}

We vary the time for which the AGN shines with a constant luminosity, $t_\mathrm{on}$, before we set its luminosity to zero, see Fig.~\ref{fig:ton}.
For $t_\mathrm{on}>0.1$\,Myr the dynamics does not change, or phrased differently: for our fiducial galaxy model it does not make a difference whether the AGN is active for $0.1$ or $10$\,Myr. In addition, this illustrates that the most significant part of the momentum transfer happens in the first $\sim10^5$\,yr. If the AGN shines only for $t_\mathrm{on}\lesssim 10^3$\,yr it cannot eject any gas out of the galaxy. The mass outflow rate increases with the lifetime as the AGN-driven wind sweeps up more ISM and reaches its peak value $\sim30$\,kyr after the AGN starts to shine. At this time, the asymmetry of the outflow starts to develop, the pressure-driving in the disc plane is less efficient because the internal energy can escape more easily perpendicularly to the disc plane, and the shock in the disc plane slows down. For $t_\mathrm{on}>30$\,kyr the mass outflow rate is independent of the lifetime. Typical AGN lifetimes are of the order $0.1-10$\,Myr with variations of the accretion rate on time-scales as short as 1000\,yr \citep{park11,sugimura16,negri17,eilers17}.


\subsubsection{Redshift}
\begin{figure}
\centering
\includegraphics[width=0.47\textwidth]{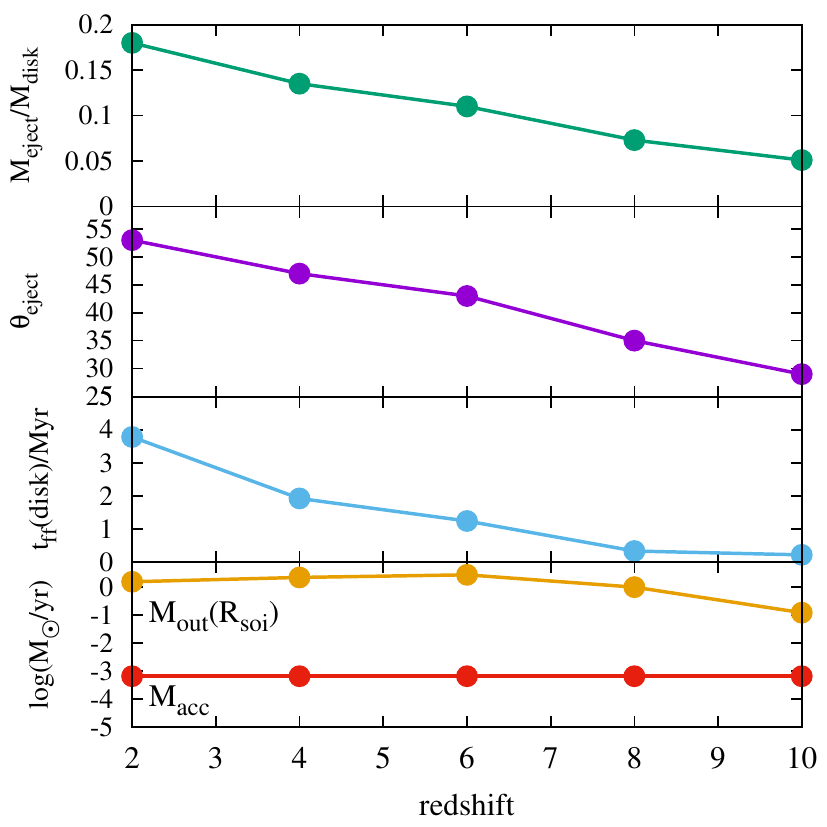}
\caption{Dependence of different quantities on  redshift. The decreasing efficiency with increasing redshift is due to the deeper gravitational potential at high redshift, because of the smaller virial radius.}
\label{fig:z}
\end{figure}

In our fiducial model we focus on low-mass galaxies at $z=6$, but our 2D approach can also be extrapolated to galaxies at other redshift and we show the redshift dependence of the AGN-driven outflow in Fig.~\ref{fig:z}.
The AGN-driven outflow is generally less efficient at higher redshift due to the deeper gravitational potential; i.e., a halo at higher redshift has a smaller virial radius for the same $M_h$, due to the higher mean cosmic density. This trend would be amplified by a smaller gas mass fraction at lower redshift, which we do not include in our model: with cosmic time, gas is converted into stars and the remaining ISM exerts a weaker resistance against the outflow, which therefore can propagate further.

\subsubsection{Initial wind velocity}
\label{sec:vinpara}
\begin{figure}
\centering
\includegraphics[width=0.47\textwidth]{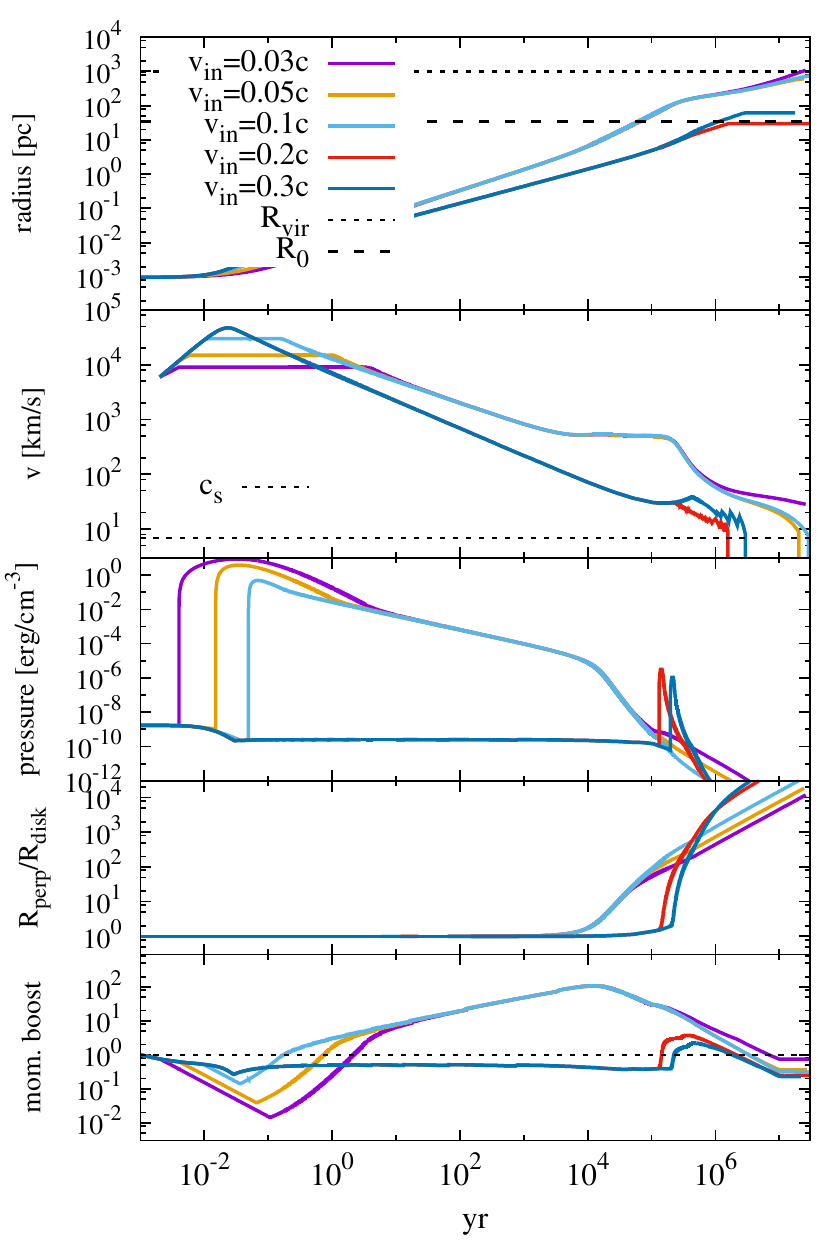}
\caption{Time evolution of the AGN-driven outflow for different values of the initial wind velocity for the direction $\theta = 45^{\circ}$. For values above our fiducial parameter of $v_\mathrm{in}=0.1c$ the transition to energy-driving occurs at much larger radii (red, dark blue). Also, the evolution of the asymmetry of the outflow is remarkable: while the shock is more spherical in the momentum-driven phase, it becomes more asymmetric after transition to energy-driving than the outflow that is energy-driven from early on (see text).}
\label{fig:vin}
\end{figure}

The initial wind velocity is crucial, because it is the least well constrained parameter and directly influences the inner wind velocity, hence the Compton cooling time, and consequently the transition from momentum- to energy-driving (Figure~\ref{fig:vin}).
For $v_\mathrm{in} \leq 0.1c$ the choice of the initial wind velocity does not affect the long-term evolution and the dynamics on galactic scales. However, for values slightly larger than our fiducial value $v_\mathrm{in} = 0.1c$, the transition to energy-driving in the disc occurs at significantly later times (see Section \ref{sec:late} for a more detailed discussion).

We can observe another interesting feature: the later the transition from momentum- to energy-driving occurs, the more asymmetric the outflow will be in the late phase ($R_\mathrm{perp}/R_\mathrm{disc}$ at $t \gtrsim 10^6$\,yr). Except for the tiny contribution of the shell self-gravity, the momentum-driven expansion in one direction does not depend on the outflow dynamics in another direction. This is in contrast to the energy-driven case, where the accelerating pressure depends on the volume enclosed by the shock front, which directly couples the outflow dynamics in different directions. Consequently, the energy-driven outflow develops an asymmetry once it encounters a path of least resistance, which makes the pressure fall and the acceleration in the disc plane smaller. In contrast, the momentum-driven outflow starts more spherical, even at larger radii, where the asymmetric density profile allows the outflow to propagate into a low-density directions. However, when the outflow now becomes energy-driven, with a higher momentum boost, the pressure does not continue to accelerate the outflow in the disc plane, because there is already a very well defined path of least resistance in the perpendicular direction. In addition, the volume enclosed by the shock front in the moment of the late transition is smaller than the volume of an outflow with an early transition to energy-driving at the same time. Hence, the pressure rises above the value of the pressure of the outflow with an early transition, compare $P(t\approx10^5\,\mathrm{yr})$.

At the same time we can see from the time evolution of the momentum boost that this higher pressure (at $\sim 10^5\,\mathrm{yr}$) does not drive the acceleration in all directions to the same degree. It rather boosts the expansion perpendicular to the disc plane, for which the momentum-driven outflow has already paved a path of least resistance. We confirm that the momentum boost depends on the confinement time, i.e. the period over which the outflow is enclosed by gas of roughly the same density.

\subsubsection{AGN luminosity}
\begin{figure}
\centering
\includegraphics[width=0.47\textwidth]{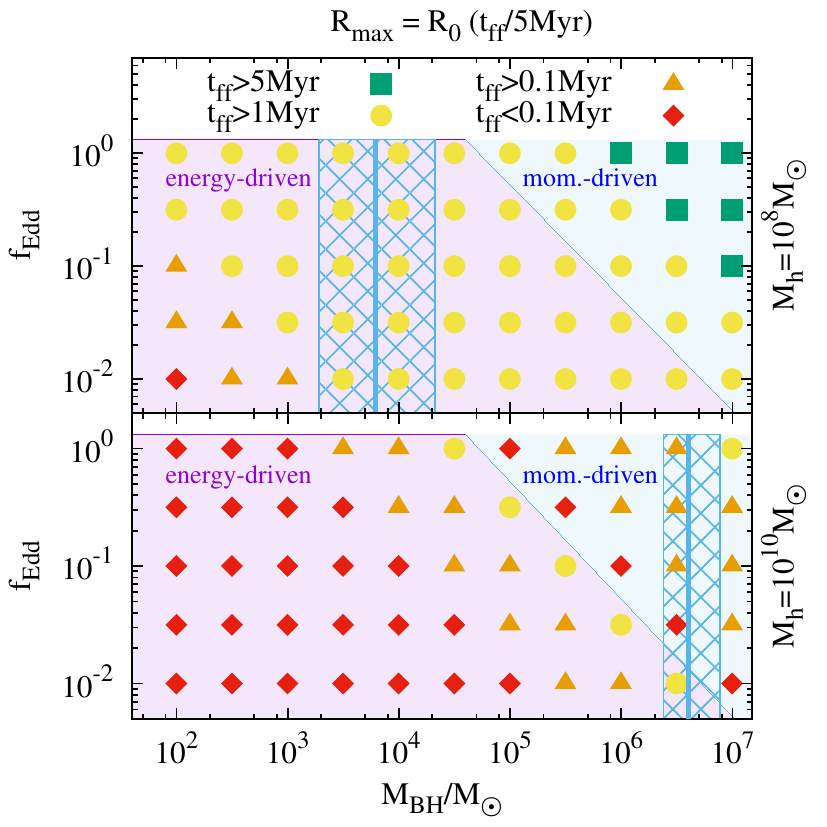}
\caption{We compare the outflow, quantified by $t_\mathrm{ff}$ in the disc plane, for different combinations of the Eddington ratio and the BH mass for two halo masses: $10^8\Msun$ (top panel) and $10^{10}\Msun$ (bottom panel). The blue vertical lines and hashed regions indicate the typical $M_\mathrm{BH}$ for these halo masses, based on the $M$--$\sigma$ relation \citep{gueltekin09}. The outflow dynamics depends only on the product $f_\mathrm{Edd}M_\mathrm{BH}$ and we clearly see two ranges of energy- and momentum-driven outflows, separated by a luminosity of $\sim 10^{43}\,\mathrm{erg}\,\mathrm{s}^{-1}$, as we will discuss in Sec. \ref{sec:late}. The same AGN in a two orders of magnitude more massive halo generates fallback times of the gas that are about one order of magnitude shorter.}
\label{fig:edd}
\end{figure}

We have verified that the outflow dynamics on all relevant scales is determined solely by the AGN luminosity, i.e. the product of Eddington ratio and BH mass and not their individual values, see Fig.~\ref{fig:edd}.
A small BH with a high Eddington ratio is therefore equivalent to a more massive BH with a smaller Eddington ratio, keeping in mind that high Eddington ratios are much less abundant then low or intermediate values \citep{habouzit17,weigel17}. The indicated freefall times yield the time that gas needs at least to fall back towards the centre once the AGN is off. It scales linearly with the radius out to which the gas is pushed in the disc plane with $t_\mathrm{ff}=5\,\mathrm{Myr}(R/R_0)$.  In Sec. \ref{sec:late} we discuss the origin of the transition to momentum-driving at higher AGN luminosities.

\begin{figure}
\centering
\includegraphics[width=0.47\textwidth]{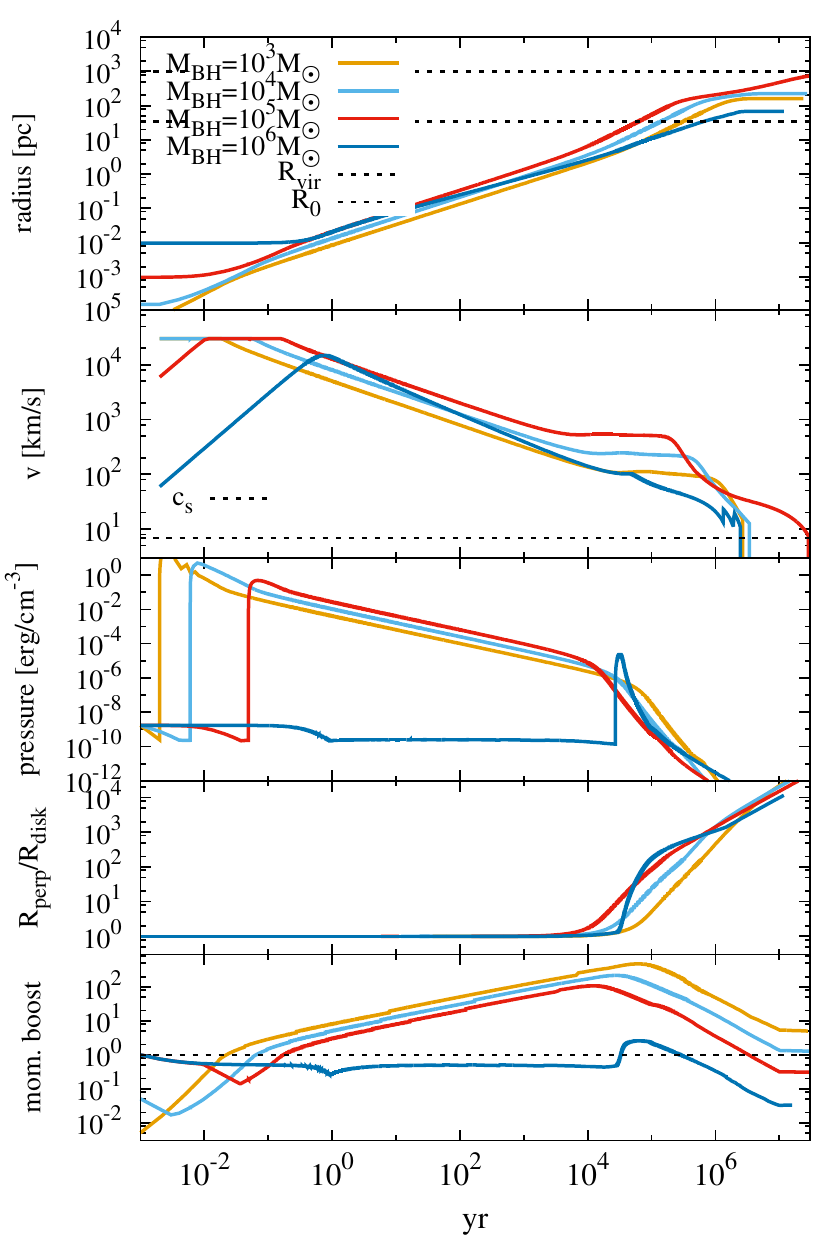}
\caption{Time evolution of the AGN-driven outflow for different BH masses, i.e. different AGN luminosities for a fixed  $f_\mathrm{Edd}=0.3$ with $M_h=10^8\Msun$. The models with $M_\mathrm{BH}\leq 10^5\Msun$ experience an early transition from momentum- to energy-driving (see sudden jump in the pressure) and the model with $M_\mathrm{BH} = 10^6\Msun$ experiences a later transition (see also section \ref{sec:late} for more details). The radius and velocity represent the evolution at $45^{\circ}$.}
\label{fig:MBH}
\end{figure}

\begin{figure}
\centering
\includegraphics[width=0.47\textwidth]{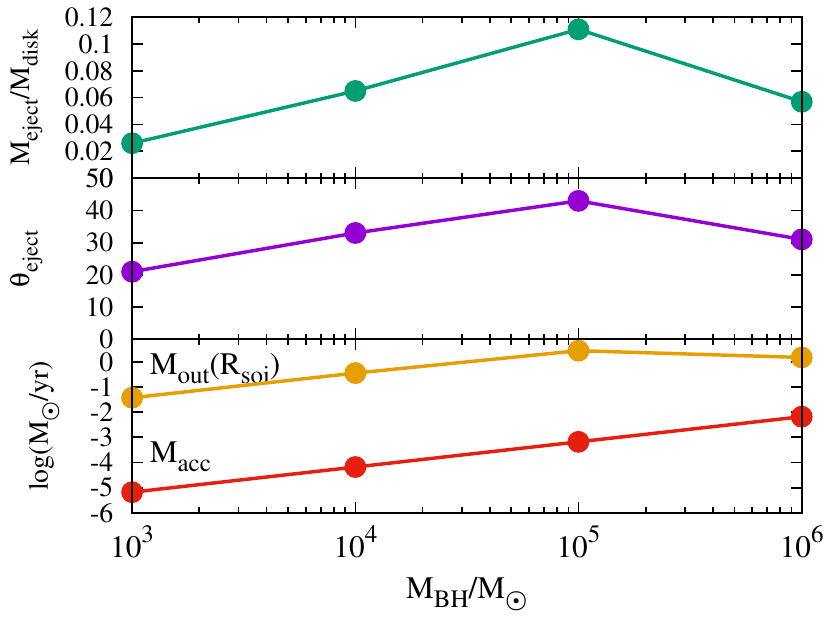}
\caption{Dependence of the gas ejection efficiency and the mass outflow rate on the BH mass for $M_h=10^8\Msun$ and $f_\mathrm{Edd}=0.3$. The ability to eject gas from the galaxy has a local peak at $M_\mathrm{BH}\approx 10^5\Msun$, because for higher BH masses, i.e. AGN luminosities, the outflow is momentum-driven in the disc (see Sec. \ref{sec:late}).}
\label{fig:MBH2}
\end{figure}

We compare the models for different BH masses, at the same halo mass, fixing the Eddington ratio at $f_\mathrm{Edd}=0.3$, in Fig.~\ref{fig:MBH} and \ref{fig:MBH2}.
We also observe an interesting trend for the momentum boost, where we observe peak values of $\gtrsim 100$, with the lowest mass black BH generating the highest momentum boosts. Phrased differently, a lower mass BH is more efficient in converting its momentum and energy input into momentum of the outflow. For all considered BH masses, the AGN can drive outflows perpendicular to the disc beyond $R_\mathrm{vir}$, but only for $M_\mathrm{BH} \gtrsim 10^5\Msun$ the AGN can eject gas at $\theta \geq 45^{\circ}$, whereas the wind powered by less massive BHs is not strong enough and it becomes subsonic before reaching $R_\mathrm{vir}$ in the diagonal direction.

The dependence of the outflow on the Eddington ratio for different BH masses, at fixed halo mass, is shown in Fig.~\ref{fig:edd}. For small BHs ($M_\mathrm{BH} \leqslant 10^3\Msun$) outflows cannot push the gas beyond disc radius even for $f_\mathrm{Edd}=1$.  A large range of the parameter space allows for outflows to affect gas in the whole disc (yellow and green symbols in the upper panel of Fig.~\ref{fig:edd}), possibly regulating the time over which an AGN has to remain off before new gas flows in to feed a new accretion episode as well as star formation in the disc.

To push gas beyond $R_0$, which is still only $\sim 3\%$ of the virial radius, a BH should be as massive as 1\% of the halo mass and have $f_\mathrm{Edd}=1$, or 10\% of the halo mass and have $f_\mathrm{Edd}>0.1$. Note that these ratios between BH and halo mass are very large. The virial velocity of a halo, i.e. the circular velocity at the virial radius, is given by $V_\mathrm{vir} = 200\,\mathrm{km}\,\mathrm{s}^{-1} (M_h/10^{11}\Msun)^{1/3}$ at $z=6$ \citep{ferrarese02,2002ApJ...581..886W,volonteri11}. Assuming that $\sigma = V_\mathrm{vir}/\sqrt{3}$ and the $z=0$ scaling between BH mass and $\sigma$ \citep{gueltekin09} yields
\begin{equation}
M_\mathrm{BH} = 6000\Msun \left( \frac{M_h}{10^8\Msun} \right) ^{1.41},
\end{equation}
for the typical BH mass in a halo of mass $M_h$.

Simulations suggest that to correctly reproduce the luminosity function of AGN only a small fraction ($\lesssim 25\%$) of high-redshift BHs in galaxies with stellar mass $<10^{10} \Msun$ is expected to have Eddington ratios above $0.01$, with the majority of BHs accreting at $f_\mathrm{Edd}<10^{-4}$ \citep{habouzit17}. For these more realistic parameter combinations, with BHs on the $M$-$\sigma$ relation and small Eddington ratios, we expect only short fallback times of the order $\lesssim 1$\,Myr. Therefore, BHs can grow almost continuously and the accretion phases are interrupted by comparably short episodes of AGN feedback: a $\sim 10^4\Msun$ BH in a $10^8\Msun$ halo has a recovery time of $\sim 1$\,Myr, indicating that accretion is not interrupted for long. A $\sim 10^6\Msun$ BH in a $10^{10}\Msun$ halo creates even shorter recovery times, making BH growth easier in higher mass systems.

The role of AGN outflows in driving large amounts of gas out of halos appears limited, if they are the only source of feedback. A viable possibility is that AGN feedback needs SN feedback as a precursor. \cite{costa14} and \cite{prieto17} show that AGN feedback is inefficient without the aid of SN feedback: SN winds heat the cold gas in the galaxy, creating a rarefied environment where energy injection from AGN feedback can easily accelerate the gas. In the next section we discuss how an outflow can still regulate the duty cycle of BHs.

In the rest of the paper we do not vary the BH mass and the Eddington ratio independently, but keep the Eddington ratio at $f_\mathrm{Edd}=0.3$ and vary the BH mass and consequently the AGN luminosity. The results for a different Eddington ratio can be rescaled accordingly.


\subsubsection{Halo mass}
\begin{figure}
\centering
\includegraphics[width=0.47\textwidth]{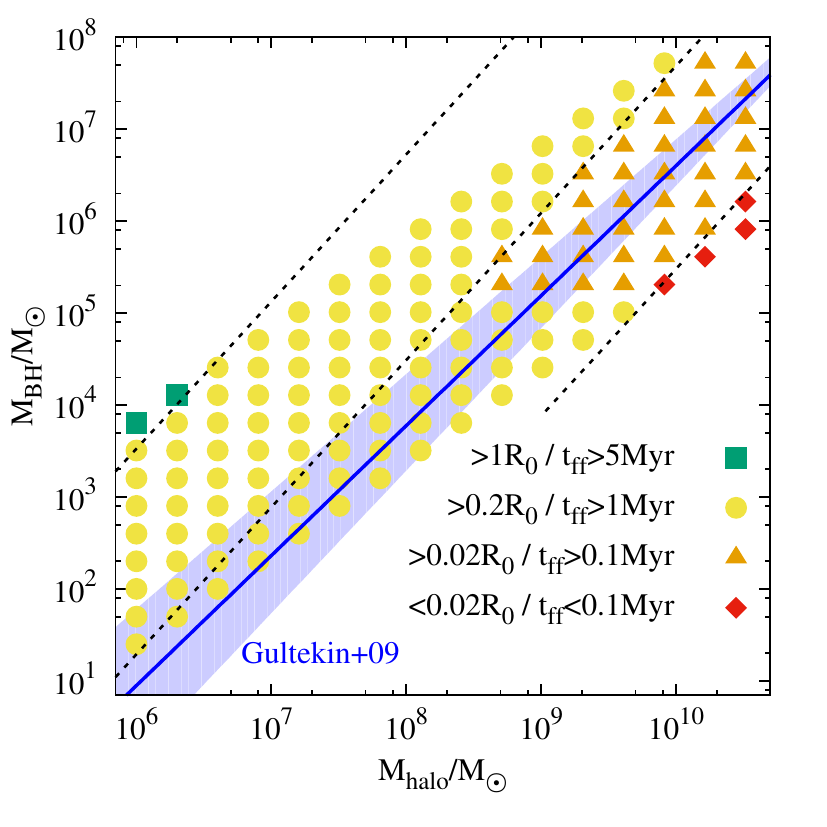}
\caption{Efficiency of ejecting gas and preventing further accretion for different combinations of the BH and halo mass, quantified by the maximum radius the outflow reaches in the disc plane. The black dashed lines represent $M_\mathrm{BH} \propto M_h^{1.6}$, illustrating thresholds of constant maximal radii. The blue line and shaded area show the expected BH masses and their error margins \citep{gueltekin09}. Note that this relation is valid for local and more massive galaxies and should rather guide the eye and indicate typical host environments for a given BH mass than be used for a quantitative comparison.}
\label{fig:MhMbh}
\end{figure}

We compare the outflow efficiency for different combinations of the halo and BH mass in Fig.~\ref{fig:MhMbh}.
As physically expected, the outflow becomes generally more powerful with a higher AGN luminosity and lower halo mass. However, there is a discontinuity of this trend at $M_\mathrm{BH}\approx 10^5\Msun$, which we will discuss in more detail in the next section. We further note that contours of equal efficiency are steeper than linear in the log($M_h$)-log($M_\mathrm{BH}$) plane. Even for the most extreme scenario of a $\sim 10^4\Msun$ BH accreting at 30\% Eddington in a $10^6\Msun$ halo, the outflow hardly pushes the gas beyond $R_0$ in the disc midplane. This indicates that the gas recovery times after AGN-driven winds are very short, of the order $\sim 1$\,Myr.


\subsection{Momentum-driven outflows for AGN luminosities $\mathbf{>10^{43}}$\,erg/s}
\label{sec:late}
\begin{figure}
\centering
\includegraphics[width=0.47\textwidth]{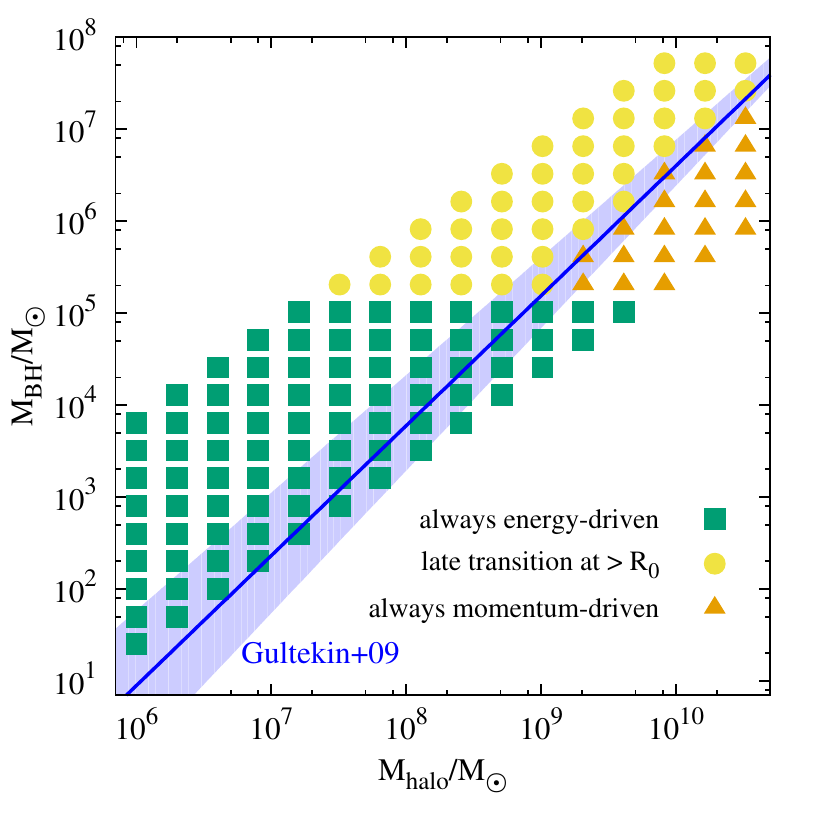}
\caption{Parameter study of the transition radius in the disc from momentum- to energy-driven outflows at $30\%$ Eddington. We can identify three different regimes: the transition occurs at very small radii close to $R_\mathrm{min}$ (green), the transition occurs at larger radii $\gtrsim R_0$ (yellow), or there is no transition at all and the outflow is always momentum-driven (orange). The last case also includes scenarios where the transition occurs at $>R_\mathrm{vir}$, which is irrelevant for our study.}
\label{fig:trans}
\end{figure}

\begin{figure}
\centering
\includegraphics[width=0.47\textwidth]{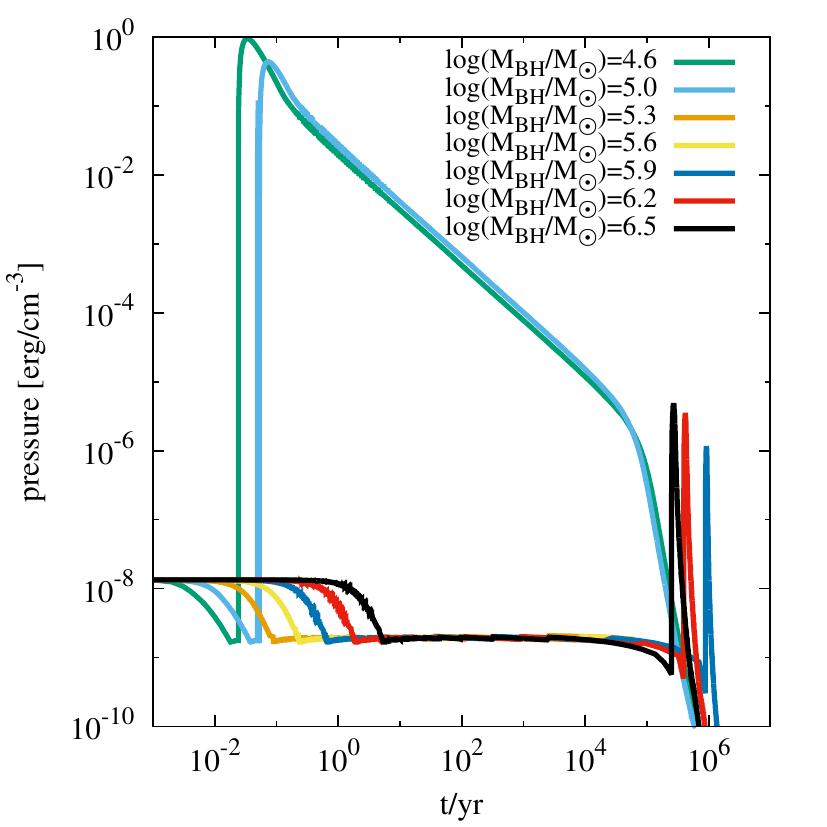}
\caption{Time evolution of the pressure in a halo with a mass of $2 \times 10^{9}\Msun$ for BH masses spanning 2 orders of magnitude. This represents a vertical slice in Fig.~\ref{fig:trans}. The transition to an energy-driven outflow occurs at different times or not at all, depending on the BH mass, i.e. on the AGN luminosity.}
\label{fig:tpMh1e10}
\end{figure}

\begin{figure}
\centering
\includegraphics[width=0.47\textwidth]{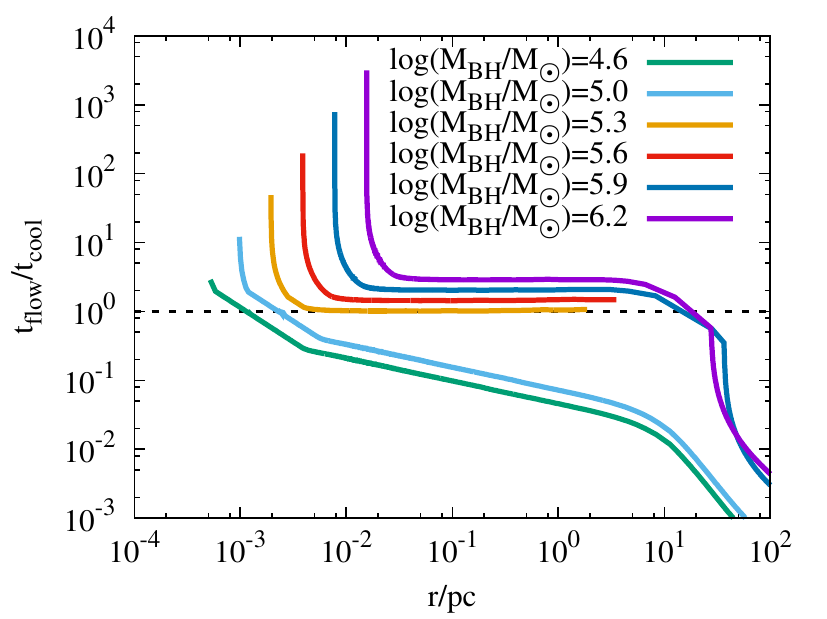}
\caption{For a fixed halo mass of $M_h=2\times 10^{9}\Msun$ we plot the ratio of the flow and the cooling time of the AGN-driven outflow as a function of radius (position of the shell) for different BH masses in the disc plane. All outflows start momentum-driven, but if the flow time becomes shorter than the cooling time, Compton cooling can no longer cool the shock efficiently and the shock becomes energy-driven. However, above $M_\mathrm{BH} \gtrsim 10^5\Msun$ the cooling time is always shorter than the flow time in the disc and both have the same scaling with radius of $t \propto r^2$. For even higher AGN luminosities, a late transition to energy-driving occurs on galactic scales.}
\label{fig:times2}
\end{figure}

We see in Fig.~\ref{fig:trans} that the transition from an initially momentum-driven to an energy-driven outflow occurs at different radii or not at all depending on the halo and BH mass for a fixed Eddington ratio.
The nature of this transition can be better seen in Fig.~\ref{fig:tpMh1e10}, where we show the time evolution of the pressure in a halo with $M_h=2\times 10^{9}\Msun$ and an outflow powered by BHs of different mass.
For $M_\mathrm{BH} \leq 10^5\Msun$ the transition to energy-driving occurs early (note that this is not our fiducial galaxy model, for which the transition occurs already at $t<0.1$\,yr). For BH masses of the order $10^5-10^6\Msun$ the outflow is always momentum-driven and for even higher BH masses of $\gtrsim 10^6\Msun$ we see a late transition to energy-driving (small peak in the red curve around $t\approx 10^6$\,yr). This trend can also be seen from a comparison of the relevant time-scales in Fig.~\ref{fig:times2}.
The three different regimes can be understood as follows: for a low BH mass (i.e. AGN luminosity) we see an early transition to energy-driving, because of the initial smaller acceleration by the AGN (green, light-blue). For significantly higher AGN luminosities (dark-blue, purple) the outflows remains momentum-driven and reaccelerates beyond $\sim 5$\,pc (compare Fig.~\ref{fig:velMap}), which decreases the flow-time and makes the outflow energy-driven. The intermediate cases (orange, red) also remain momentum-driven, but become subsonic before they reach $\sim 5$\,pc.

We can understand these transitions at systematically different radii based on simple analytical arguments. The Compton cooling time as a function of radius $r$ is given by Eq. \ref{eq:compton2} or for typical parameters:
\begin{align}
&t_\mathrm{compton} = \nonumber \\
&1.4\,\mathrm{kyr} \left( \frac{f_\mathrm{Edd}}{0.3} \right) ^{-1} \left( \frac{M_\mathrm{BH}}{10^6\Msun} \right) ^{-1} \left( \frac{v_\mathrm{in}}{0.1 c} \right) ^{-2} \left( \frac{r}{1\,\mathrm{pc}} \right) ^{2}.
\end{align}
To derive the relevant flow time, we have to make several simplifying assumptions. If the transition to energy-driving occurs, it occurs at $<1$\,pc (compare Fig.~\ref{fig:times2}), which is within the sphere of influence and the density in this inner part can be assumed to be constant.
Furthermore we neglect the gravity of the BH. This assumption might seem to be in contradiction to being within the BH's sphere of influence, but the gravitational force is proportional to the mass of the swept-up ISM, which is negligibly small at the radii of interest (compare Fig.~\ref{fig:acc}). The equation of motion is then given by
\begin{equation}
\frac{\mathrm{d}}{\mathrm{d}t}(M_\mathrm{shell}v)=\frac{L}{c},
\end{equation}
with the shell's velocity $v=\dot{r}$ and the shell mass $M_\mathrm{shell} = 4/3 \pi r^3 \rho _0$. The central gas density $\rho _0$ can be derived from the exponential disc profile as
\begin{equation}
\rho _0 = \frac{\Sigma _0}{2 H_0} = \frac{G}{2\pi c_s^2}\frac{(m_d M_\mathrm{halo})^2}{(\lambda R_\mathrm{vir})^4},
\end{equation}
where $H_0=H(R=0)$ is the disc scale height in the centre. The differential equation can be solved by a function of the form
\begin{align*}
r(t)&=r_0 \left( \frac{t}{t_0} \right) ^{0.5}\\
v(t)&=\frac{r_0}{2} (t_0t)^{-0.5},
\end{align*}
which yields directly the same scaling with radius for the flow time as for the cooling time:
\begin{equation}
t_\mathrm{flow} = \frac{r}{v} = 2 t \propto r^2.
\end{equation}
We note again that this result is only valid for the inner part of the disc, where the density can be assumed to be constant and the mass of the swept-up gas is still negligible for the gravity. For the flow time as a function of radius we obtain
\begin{equation}
t_\mathrm{flow} = \sqrt{\frac{8 \pi c \rho_0}{3L}} r^2 \propto m_d \lambda ^{-2} (1+z)^{3/2} f_\mathrm{Edd}^{-1/2} M_\mathrm{BH}^{-1/2} R^2.
\end{equation}
Or expressed with fiducial values and fixed $m_d=\lambda=0.05$ at $z=6$:
\begin{equation}
t_\mathrm{flow} = 2.2\,\mathrm{kyr} \left( \frac{f_\mathrm{Edd}}{0.3} \right)^{-1/2} \left( \frac{M_\mathrm{BH}}{10^6\Msun} \right)^{-1/2} \left( \frac{r}{1\,\mathrm{pc}} \right)^{2}.
\end{equation}
Setting it equal to the cooling time we can conclude that the outflow is always momentum-driven if the BH and Eddington ratio fulfil
\begin{equation}
\frac{f_\mathrm{Edd}}{0.3}  \frac{M_\mathrm{BH,crit}}{4 \times 10^5\Msun} > \left( \frac{v_\mathrm{in}}{0.1 c} \right) ^{-4}  \left( \frac{1+z}{7} \right) ^{-3},
\label{eq:MBHtrans}
\end{equation}
or expressed via the AGN luminosity
\begin{equation}
L_\mathrm{AGN} > 10^{43}\,\mathrm{erg}\,\mathrm{s}^{-1} \left( \frac{v_\mathrm{in}}{0.1 c} \right) ^{-4}  \left( \frac{1+z}{7} \right) ^{-3}.
\end{equation}
Note the independence on the halo mass and the strong dependence on the not well-constrained initial velocity of the inner wind $v_\mathrm{in}$. Uncertainties in this parameter can change the resulting transition luminosity by up to three orders of magnitude. The analytically derived BH transition mass, $4\times 10^{5}\Msun$, is close to the one obtained in the 2D simulation of $\sim~2\times 10^{5}\Msun$. The small difference is related to the necessary approximations in order to analytically solve the equation of motion. The strong redshift dependence indicates that the transition mass above which a BH can no longer drive an efficient energy-driven wind in the disc is higher at low redshift, caused by the redshift dependence of the central density.

The parameter dependences can be explained as follows: a higher Eddington ratio $f_\mathrm{Edd}$ yields more photons by the AGN, which boost the cooling via inverse Compton scattering. At lower redshift, the gravitational potential is shallower for the same halo mass, which makes the shell velocity higher, the flow time smaller, and hence requires more efficient cooling to sustain a momentum-driven outflow. The initial wind velocity is directly proportional to the post-shock temperature (Eq. \ref{eq:Tpost}) and hence defines the internal energy of the shocked wind.

Momentum-driven outflows around the threshold luminosity $L_\mathrm{AGN} \approx 10^{43}\,\mathrm{erg}\,\mathrm{s}^{-1}$ or slightly above generate a smaller momentum boost (Fig.~\ref{fig:MBH}), are less powerful in ejecting gas out of the galaxy (Fig.~\ref{fig:MBH2}), and drive the gas in the disc plane to smaller maximal radii (Fig.~\ref{fig:edd}, \ref{fig:MhMbh}). More luminous AGNs with $M_\mathrm{BH}\approx 10^7\Msun$ and $f_\mathrm{Edd}=1.0$ in $10^8\Msun$ halos are the only systems that create recovery times of the gas in excess of 5\,Myr (Fig.~\ref{fig:edd}) via momentum-driving. However, such a combination of Eddington ratio, BH and halo mass is very unlikely.

\subsection{Gas ejection perpendicular to the disc}
\label{sec:perp}
\begin{figure*}
\centering
\includegraphics[width=0.99\textwidth]{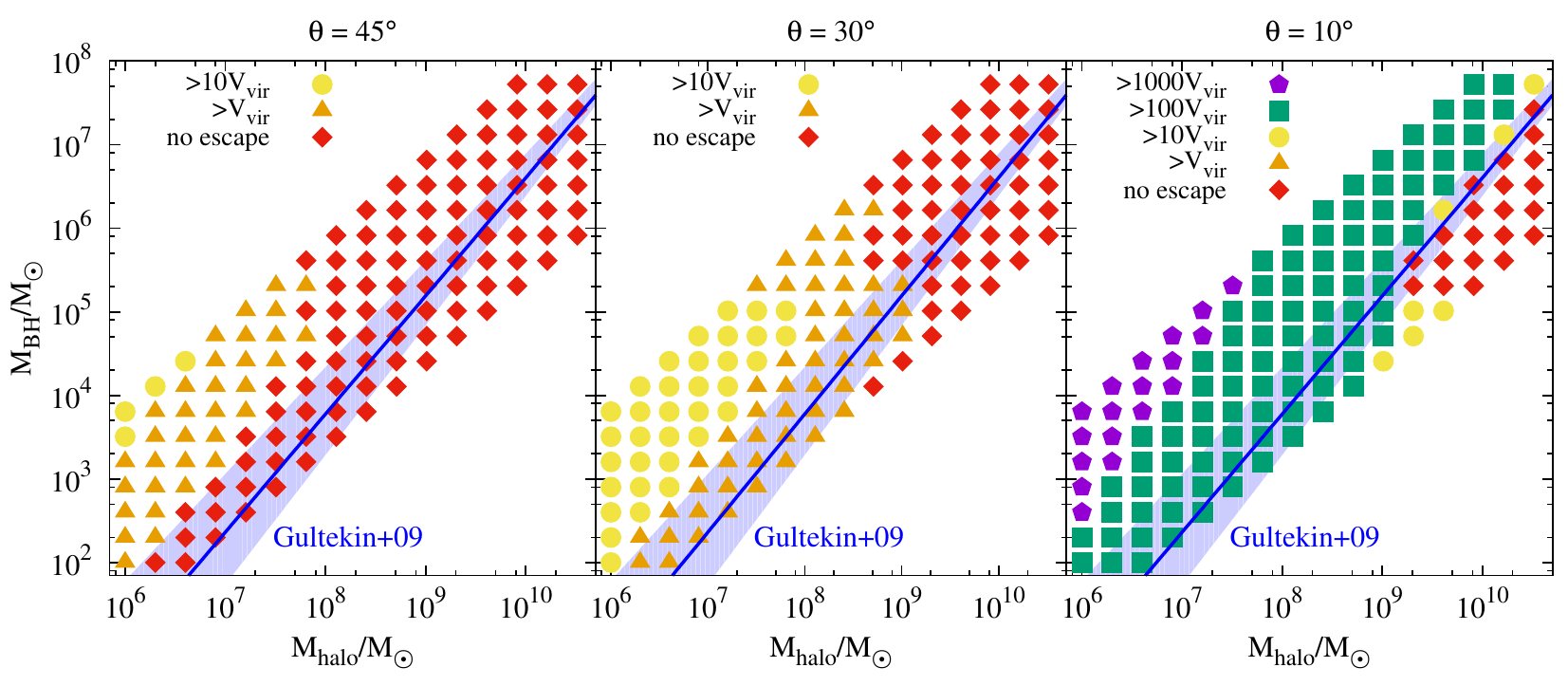}
\caption{Outflow velocities at the virial radius, normalized to the virial velocity for different angles with respect to the disc normal. All models are with $f_\mathrm{Edd}=0.3$ and the general trend is that more massive BHs in less massive galaxies create higher outflow velocities, close to the disc normal even in excess of $10^4\,\mathrm{km}\,\mathrm{s}^{-1}$. However, there is also a wide range of models that cannot eject gas at $\theta > 45^{\circ}$ beyond the virial radius.}
\label{fig:esc}
\end{figure*}

So far, we have focused on the gas dynamics in the disc plane and have seen that AGN-driven winds in 2D push the gas in the disc plane at most to about the scale radius $R_0$. If we instead focus on the gas dynamics and outflow velocities at higher inclination, we can study the wind on galactic scales and relate it to observations of gas dynamics in high-redshift galaxies. In Fig.~\ref{fig:esc}, we compare the outflow velocities at the virial radius for different angles with respect to the disc.
For better comparison, we normalize the velocities to the virial velocity of the corresponding halos,
\begin{equation}
V_\mathrm{vir} = 20\,\mathrm{km}\,\mathrm{s}^{-1} \left( \frac{M_h}{10^8\Msun} \right)^{1/3}.
\end{equation}
A low halo mass and a high AGN luminosity generate outflows with higher velocities, as naively expected. As before, this general trend is interrupted by the transition between energy- and momentum-driving around a BH mass of $10^5\Msun$. Above this threshold, the outflow remains energy-driven at high inclination, but turns momentum-driven in the disc and a part of the post-shock thermal energy is radiated away. This reduces the pressure and therefore the adiabatic acceleration in all directions.

There is a large parameter range, for which gas at $\theta \geq 45^{\circ}$ cannot escape the halo (red diamonds). In contrast, close to the disc normal, gas is energy-driven and ejected with high velocities of $>10^4\,\mathrm{km}\,\mathrm{s}^{-1}$. AGN-driven outflows eject gas out of the galaxy and the direction-dependent velocities at $R_\mathrm{vir}$ range from $\lesssim 100\,\mathrm{km}\,\mathrm{s}^{-1}$ to mildly relativistic velocities close to the disc normal in agreement with observations \citep{chartas02,pounds03,cappi06,feruglio10,rupke11,aalto12b,gofford13,cicone14}. The subset of models that can not even eject gas at $\theta = 10^{\circ}$ (red diamonds in the right panel) are the models where the outflow is always momentum-driven in the disc (compare Fig.~\ref{fig:trans}). However, one should keep in mind that these results were derived for an Eddington ratio of $30\%$ and higher values increase gas ejection (only the product of Eddington ratio and BH mass enters in the equation of motion and results can be rescaled accordingly).

\subsection{Comparison to spherical case}
\begin{figure}
\centering
\includegraphics[width=0.47\textwidth]{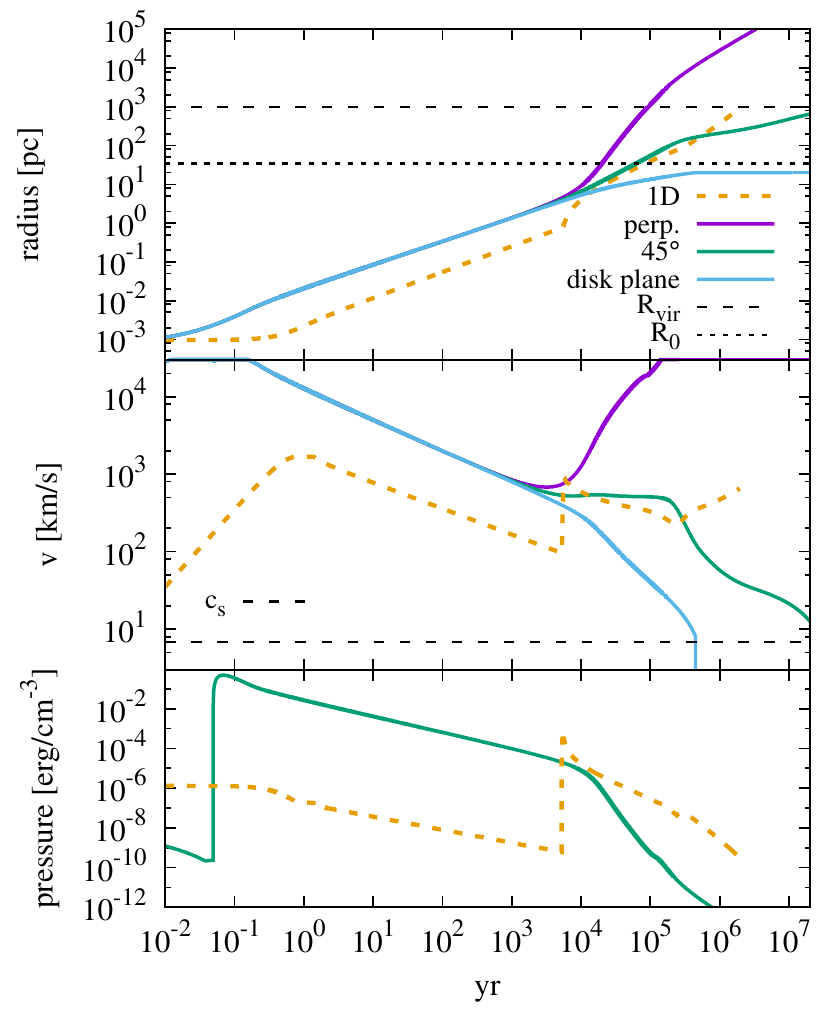}
\caption{Same as Fig.~\ref{fig:trvp}, but with the additional solution of a 1D model (orange, dotted), where we distribute the disc gas mass following a Hernquist profile. The transition to momentum-driving occurs at later time and larger radius, which is partially caused by the different radial density profiles for the Hernquist profile in 1D and an exponential disc in 2D. In the spherical model, all gas is ejected beyond $R_\mathrm{vir}$.}
\label{fig:trvp1D}
\end{figure}

In this section we highlight the importance of a 2D approach by comparing it directly to the solutions obtained in a 1D model. We assume that the gas in the 1D model follows the distribution of the DM, which we describe by a Hernquist profile (Eq. \ref{eq:HernquistGas}). We first compare the general dynamics in our fiducial model in Fig.~\ref{fig:trvp1D}.
The 1D outflow is momentum-driven for a longer time and therefore starts with smaller velocities. However, once the 1D outflow becomes energy-driven, it ejects all the gas out of the virial radius, whereas the 2D model can only eject a smaller fraction of the total gas. 
\begin{figure}
\centering
\includegraphics[width=0.47\textwidth]{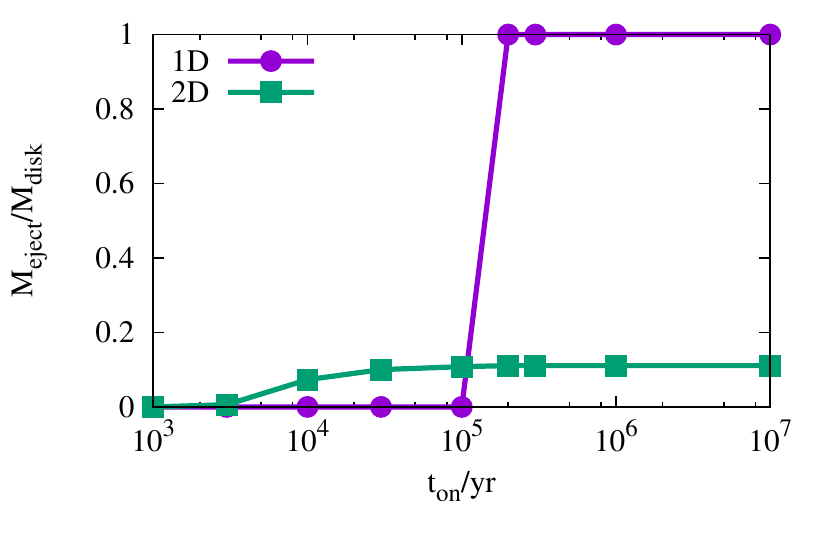}
\caption{Fraction of the disc mass that is ejected by the AGN-driven wind as a function of the time over which the AGN shines with a constant luminosity (fiducial model with $f_\mathrm{Edd}=0.3$). In the 1D model either no or all the gas is ejected out of the virial radius and this transition occurs at $t_\mathrm{on}\approx 10^5$\,yr. In our 2D model we see a gradual rise of the ejected mass until it reaches the final value of $\sim 10\%$ at $t_\mathrm{on}\gtrsim 10^5$\,yr. This demonstrates the strength and importance of a 2D treatment of AGN-driven feedback, compared to a simplistic 1D model.}
\label{fig:Meject}
\end{figure}
We further quantify this effect in Fig.~\ref{fig:Meject}. In a spherical model the ejected mass is either 0\% for $t_\mathrm{on}\leq 10^5$\,yr or 100\% for $t_\mathrm{on}> 10^5$\,yr, whereas in our 2D model the ejected mass increases gradually with $t_\mathrm{on}$, and reaches saturation at $11\%$ for $t_\mathrm{on}\gtrsim 10^5$\,yr. In this regime, the 1D model overestimates the efficiency of an AGN-driven outflow by one order of magnitude in our fiducial galaxy model. It is difficult to further extrapolate these results to 3D. If we assume the same disc-like geometry in 3D, the ejection fractions and fall-back times may not change much. However, a 3D model allows for a more accurate treatment of the inhomogeneous ISM, which could increases or decreases the overall outflow efficiency, as we discuss in Sec. \ref{sec:cav}.

\begin{figure}
\centering
\includegraphics[width=0.47\textwidth]{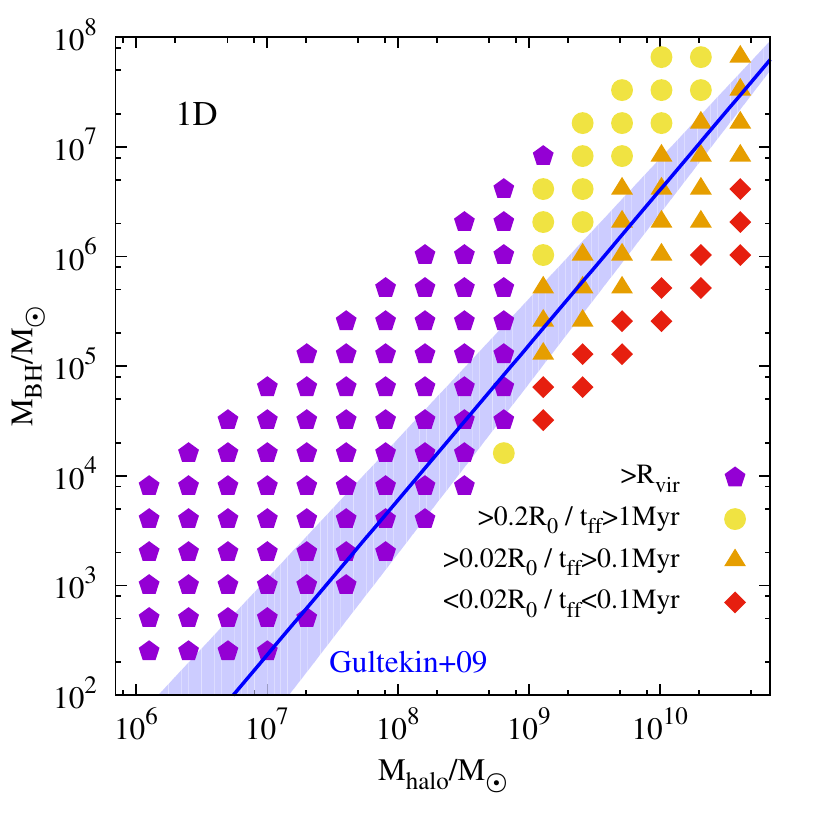}
\caption{Maximum radius the AGN-driven wind reaches in spherical symmetry for different combinations of the BH and halo mass (compare to the 2D results in Fig.~\ref{fig:MhMbh}). The purple pentagons in this plot indicate complete gas ejection out of the galaxy.}
\label{fig:MhMbh1D}
\end{figure}
In Fig.~\ref{fig:MhMbh1D} we plot the efficiency of the 1D model in ejecting gas out to a certain radius. To quantify the efficiency of the outflow in the 1D Hernquist profile we apply the same scaling radius, $R_0$, as for 2D exponential disc (Eq. \ref{eq:ScalingRadius}) to check if the outflow can reach at least this radius. The enclosed gas mass within $R_0$ in both scenarios is roughly the same with less than a factor two difference. There is a large parameter range for which 1D models can eject all the gas out of a galaxy. This range is of special interest, because it encloses halo masses of $\sim10^7-10^9\Msun$, which are typical masses of dwarf galaxies at low redshift and of high-redshift galaxies, which might host the first massive BH seeds. Above a certain halo mass of $\sim 10^9\Msun$, which corresponds to a typical BH mass of $10^5\Msun$, the AGN (at 30\% Eddington) is not powerful enough to eject gas out of the galaxy.

We compare these results to \cite{park16} and \citet{pacucci15}, who use 1D radiation-hydrodynamics simulations to study the effect of radiative feedback on the growth of BHs in small halos. They show that light BHs of $\sim 100\Msun$ cannot grow efficiently via radiation-regulated accretion. \citet{pacucci15} further derive an analytical expression for this critical BH mass, which is in the range $10$--$10^6\Msun$, depending on the accretion scenario and the host properties. In our 1D comparison model, we find a comparable mass range of $M_\mathrm{BH} \approx 10^2-10^7\Msun$, where the BH can completely eject the gas out of a host halo ($M_h \lesssim 10^9\Msun$) and hence prevent rapid mass growth of the BH. For a fixed halo mass and Eddington ratio in two dimensions, a higher mass BH generates stronger outflows and ejects the gas to larger radii. This makes lower mass BHs less efficient in stopping their own gas supply and hence more susceptible for mass growth at high duty cycles.
This is the inverse trend of \citet{pacucci15} and \citet{park16}, who find low mass BHs to be more efficient in stopping their own gas supply. These differences are likely caused by different assumptions on what drives the outflow: \citet{pacucci15} and \citet{park16} study the direct influence of radiation pressure from the BH on the accretion process, while we model AGN-driven winds. For instance, the critical mass for radiation pressure to significantly change the mass accretion rate \citep{pacucci15} and reduce growth is related to the significance of the Eddington limit. The inflow rate is determined by the halo properties. If a given halo provides the same amount of gas inflow to a small BH or a large BH, it will be the smaller BH that will reach the Eddington limit first, having its growth stunted.


We note that the differences between the 1D and 2D model presented in this section might partially depend on the different scaling of the density with radius and hence on the different radii where the outflow becomes energy-driven. However, the qualitative differences between the 1D and 2D model remain still valid beyond these differences in the density profile.

\section{Discussion}
\label{sec:discussion}
We have developed a new 2D analytical model of AGN-driven outflows and present the results for various galaxies and a range of typical conditions. The additional dimension with respect to previously existing 1D models allows to account for the direction-dependent density profiles. This has significant consequences on the maximally ejected gas mass fraction, on the momentum boost, and on the condition to stop further accretion on the central BH.


\subsection{Advantage of 2D}
With a more realistic disc-profile we confirm the results of 3D hydrodynamic simulations that the outflow propagates preferentially towards low density regions. In the momentum-driven regime this is simply due to the lower column densities perpendicular to the disc and hence the lower gas masses that have to be accelerated. In the energy-driven regime, however, the evolution of the entire confined volume is relevant for the acceleration of the shock. Once the energy-driven shock has created a chimney perpendicular to the disc, the pressure and internal energy can escape and the driving in the disc plane is significantly reduced.

Naturally, we find that an AGN drives the gas out to different radii for different angles, before the shock front becomes subsonic or crosses the virial radius. In 1D models there is only one threshold AGN luminosity above which all the ISM is ejected and below which the AGN wind is not efficient enough. For none of our models the AGN can eject all the gas out of the galaxy. 

If we define the criteria to stop further accretion on to the SMBH via the minimum time that the swept-up gas needs at least to fall back to the centre, or equivalently that gas is swept out at least to a given radius in the disc plane, we find that this condition results in a proportionality of $M_\mathrm{BH} f_\mathrm{Edd} \propto M_h^{1.6}$ (Fig.~\ref{fig:MhMbh}). The power of this scaling is independent of the exact maximal radius or minimal time chosen. However, the absolute value of this threshold has to be quite small ($R_\mathrm{max}\lesssim 0.01R_\mathrm{vir}$), if we want to reproduce the observed $M$--$\sigma$ relation with our 2D analytical model of AGN feedback (blue line in Fig.~\ref{fig:MhMbh}).

For $\sigma \approx V_\mathrm{vir} \propto M_\mathrm{h}^{1/3}$ \citep{ferrarese02,loeb10,volonteri11} we find a scaling between the BH mass and the halo velocity dispersion of
\begin{equation}
M_\mathrm{BH} \propto \sigma ^{4.8},
\end{equation}
which is close to the proposed slope of $M_\mathrm{BH} \propto \sigma ^5$ for an energy-driven outflow in 1D models \citep{silk98,haehnelt98,fabian12,McQuillin13}.

We do not include SN feedback in our model, which is also expected to change this relation between the galaxy mass and the efficiency to eject gas. We compare SN-driven outflows to AGN-driven winds in more detail in \citet{dashyan17}. Moreover, our properties of the galactic disc are intrinsically linked to the DM halo by assuming that a constant fraction of the gas with a certain fraction of the angular momentum settles into the disc \citep{mo98}. Therefore, we cannot disentangle the effect of our 2D density distribution and the underlying DM halo on the M-$\sigma$ scaling with our analytical model.

Our new 2D model further allows for new interpretations of existing observations.
\citet{cicone15} present \CII~and FIR continuum observations with the Bure Interferometer of a quasar and its host galaxy at $z=6.4189$. The \CII~reveals velocities up to $1400 \mathrm{km}\,\mathrm{s}^{-1}$ and outflow with a complex morphology out to $\sim 30$\,kpc. They find no evidence for a regular rotation pattern and identify 48 individual clumps with dynamical times in the range $10^{6.6-8.0}$\,yr. They interpret the spread in outflow times as a non-constant AGN luminosity causing various outbursts. However, it could also be related to projection effects or simply different flow times in different directions with respect to the disc plane (compare Fig.~\ref{fig:self}). This supports the importance of modelling AGN outflows in 2D to allow for an outflow with different velocities in different directions.

\subsection{Outflow efficiency as a function of the AGN luminosity}
The efficiency of an outflow can be defined in different ways: either by its ability to convert input energy into momentum of the outflow (momentum boost), or by its ability to evacuate the ISM within a certain radius and preventing further gas accretion for a certain time. In this section, we use these two quantifications to compare AGN-driven outflows in the momentum- and in the energy-driven regime.

BHs of lower mass or accreting at lower Eddington ratio have a higher momentum boost: that is, they have a stronger push relative to the input momentum. The momentum loading of the outflow increases with decreasing AGN luminosity, because if the luminosity is too high, the outflow rapidly propagates towards lower density regions and paves a path of least resistance perpendicular to the disc plane. Consequently, the shocked wind can adiabatically expand in this direction without transferring its momentum to the denser gas in the disc plane. An outflow driven by a lower luminosity AGN remains confined for longer and therefore has more time to transfer its momentum to the gas. 

However, the strength of the outflow, defined as its ability to push gas out to a given radius, e.g., the disc radius $R_0$, is higher the more massive and highly accreting BHs are (Fig.~\ref{fig:edd},~\ref{fig:MhMbh}). High-luminosity AGN are more efficient in driving an outflow out to large galactic radii and hence suppress their own gas supply for a significant amount of time. For instance Fig.~\ref{fig:edd} shows that in a halo of $10^8 \Msun$ an AGN with luminosity below $<10^{40}$\,erg\,s$^{-1}$ shuts off accretion for only $<1$\,Myr, an AGN with luminosity $10^{40}$\,erg\,s$^{-1}<L_\mathrm{AGN} \lesssim 10^{43}\,\mathrm{erg}\,\mathrm{s}^{-1}$ shuts off accretion for $\sim 1$\,Myr, and an AGN with luminosity above $>10^{44}$\,erg\,s$^{-1}$ shuts off accretion for $>10$\,Myr. These AGN luminosities of log$(L_\mathrm{AGN}/\mathrm{erg}\,\mathrm{s}^{-1})=40,~43,~44$ correspond to BH masses of roughly $10^2,~10^5,~10^6\Msun$ at $f_\mathrm{Edd}=1.0$. In a more massive halo these time-scales are even shorter, since they depend on the virial radius.

For AGN luminosities $\gtrsim 10^{43}\,\mathrm{erg}\,\mathrm{s}^{-1}$, the outflow remains momentum-driven in the disc out to galactic radii. For AGN luminosities around this threshold, the momentum-driven outflow is less powerful in ejecting gas out of the galactic potential and in preventing further mass accretion in the disc plane, compared to an energy-driven outflow in the same galaxy. However, at even higher AGN luminosities, the AGN is more efficient and can prevent further gas infall for $>10$\,Myr, whereas the energy-driven outflow can only push gas in the disc plane out to distances, which correspond to a recovery time of $\sim 1$\,Myr. The main reason for this difference is that the direct momentum input acts more isotropically and works still in the higher-density disc plane, even if there is already a path of least resistance in the perpendicular direction. In contrast to the energy-driven outflow, where all the pressure can escape perpendicular and does not push the gas in the disc plane to larger radii.


\subsection{Driving mechanism}
For low AGN luminosities, the transition from momentum- to energy-driving in the disc plane occurs on very small, i.e. sub-pc, scales in our models. For higher AGN luminosities beyond $L_\mathrm{AGN} \gtrsim 10^{43}\,\mathrm{erg}\,\mathrm{s}^{-1}$, we observe that the outflow remains momentum-driven in the disc.
In all models we see a mildly relativistic, energy-driven outflow close to the disc normal. This outflow develops due to the low column density in this direction, which causes very high accelerations. The ejected mass in this cone is negligible compared to the gas mass in the disc.
Previous 1D models find that the transition from momentum- to energy-driving occurs on scales of  $\sim~~1$\,kpc \citep{ciotti97,king03,king11,zubovas12}, several orders of magnitude larger than what we find in our 2D model. In our 1D comparison model this transition occurs around $1$\,pc (Fig.~\ref{fig:trvp1D}), but our halo mass is lower than in other 1D studies.

\citet{bourne13} model the expected X-ray signature of inverse Compton cooling in a one temperature medium and with thermally decoupled electrons in the post-shock wind. They argue that current observations of AGN do not show evidence of Compton cooling from a one temperature medium and weak constraints on a possible cooling from a two-temperature medium. This observation supports the theory of energy-driven winds on galactic scales that do not radiate away their thermal internal energy. This suggests that most AGN-driven outflows are energy-driven.

We derive a characteristic AGN luminosity above which the wind remains momentum-driven in the galactic disc (Eq.~\ref{eq:MBHtrans}). For our fiducial parameters this characteristic luminosity corresponds to a BH mass of the order $\sim 10^5\Msun$, which is independent of the halo mass, but it strongly depends on the redshift and the initial wind velocity.

In this paper, we focused on galactic outflows driven by the inner disc wind of the AGN. However, the ISM could also be accelerated directly by the radiation pressure from the AGN \citep{ishibashi14,ishibashi15,thompson15,costa17}. Although this mechanism can be efficient close to the SMBH, radiation pressure can accelerate gas only up to an effective optical depth of order unity. For electron scattering this transparency radius, where $\tau \approx 1$, is \citep{king14,king15}
\begin{equation}
R_\mathrm{tr} \approx 50\,\mathrm{pc} \left( \frac{f_\mathrm{gas}}{0.17} \right) \left( \frac{\sigma}{200\,\mathrm{km}\,\mathrm{s}^{-1}} \right) ^{2},
\end{equation}
beyond which direct driving by radiation pressure from the AGN cannot accelerate galactic outflows. The inclusion of dust increases the efficiency of direct radiation pressure on the ISM, due to the higher cross-section of dust and the consequently larger transparency radius of several kpc \citep{ishibashi12,ishibashi15,king15}. The momentum input of such a radiation pressure-driven wind can be comparable to that of a momentum-driven wind, but the main difference is the frequency-dependent cross-section of dust (with a peak in the UV). Consequently, not all gas might experience the radiation pressure due to its locally low opacity or due to self-shielding by higher opacity regions. Even hybrid models might be realistic, where the inner wind does not directly emerge from the disc, but is accelerated by radiation pressure. In any case, if radiation pressure on electrons and dust drives galactic gas outflows, the accelerated ISM prefers the path of least resistance, which requires a 2D treatment.

\subsection{Inferences on quasar outflows}
The AGN-driven outflows of our fiducial model eject gas out of the galaxy and the direction-dependent velocities at $R_\mathrm{vir}$ range from $\lesssim 100\,\mathrm{km}\,\mathrm{s}^{-1}$ to mildly relativistic velocities close to the disc normal. Most observations, however, target higher-luminosity, lower-redshift AGN. Our model is not intended to simulate quasars in high mass galaxies at relatively low redshift, where the gas content is expected to be lower than the universal baryon fraction and a large stellar component is expected. We can, however, speculate on the qualitative behaviour of the outflows by studying the properties of a luminous AGN with luminosity $10^{47}\,\mathrm{erg}\,\mathrm{s}^{-1}$ in a halo with mass $\sim 10^{12.5}\Msun$ at $z\sim 2$ as a reference case \citep[e.g.,][]{2013ApJ...776..136P} with 10\% and 100\% gas fraction. 

Also in these models velocities range from $\lesssim 100\,\mathrm{km}\,\mathrm{s}^{-1}$ to almost $ 10^4\,\mathrm{km}\,\mathrm{s}^{-1}$, in agreement with observations. Gas is able to escape from $\sim$ 60\% to 10\% of the total solid angle for 10\% and 100\% gas fractions, respectively. This is the region in the halo within which the outflow is energy-driven (see Sec. \ref{sec:perp} for details). The amount of gas escaping from a given direction is, however, not a monotonic quantity. A small amount of gas is pushed at small opening angles around the normal to the disc, as the subtended gas mass is small; the gas mass increases with increasing distance from the normal, but eventually it drops, when the outflow in that direction is unable to escape, analogously to what we see in Fig.~\ref{fig:densMap} for our fiducial set of parameters. 

This appears in nice agreement with the observational results of  \cite{2007ApJ...655..735H} and \cite{2013ApJ...776..136P}, who find a low incidence of gas absorbers along the line of sight to quasars (``down-the-barrel"), and they argue that gas in this direction is photoionized, while the incidence of absorbers in the transverse direction is larger towards smaller scales, implying an increasing density and covering fraction of neutral gas towards the quasar. 

The outflow properties in terms of velocity, asymmetry, and energetics are also in good agreement with observations \citep{2009ApJ...690.1558P}. One important difference is the gas temperature. The observed gas is cool, $\sim 10^4$ K, and cannot have been shock heated recently, which disfavours AGN-driven outflows (especially energy-driven ones) as an interpretation. However, depending on the time-scales, the gas could already have cooled down, after being pushed out by the AGN wind.

\subsection{Caveats}
\label{sec:cav}
We treat the ISM as homogeneous and assume that the AGN-driven shock sweeps up all the enclosed gas. Real galaxies have a multiphase ISM with H{\sc ii} regions and dense molecular clumps, which alter this simplistic treatment. Theoretical models of the impact of AGN winds \citep{bieri17} and jets \citep{wagner12,wagner13,cielo17} on a clumpy ISM show that dense clumps can be dissolved by IR photons that can penetrate into these clumps. The efficiency of the AGN-driven wind depends on the size and filling factor of these dense clumps. A galaxy with many small isolated clouds experiences efficient cloud dispersion compared to a galaxy with fewer but bigger cloud complexes. By neglecting the realistic ISM structure, we overestimate the outflow efficiency, because a fractal ISM predefines already paths of least resistance and the momentum transfer to dense clumps is smaller compared to a homogeneous ISM. On the other hand, the low-density gas between the clumps is accelerated and ejected more easily.

The shock can also form dense cold clumps via dynamical instabilities as shown by \citet{costa15}, which we do not include in our model. The contact discontinuity is strongly Rayleigh-Taylor unstable and might give rise to the formation of clumps that form behind the shock front and decouple from the outflow. These denser entities could also be the constituents of the high-speed molecular outflows. Recently, \citet{ferrara16} model the formation of molecular clumps in AGN-driven outflows with 3D simulations. They find that clumps might form at the transition from momentum- to energy-driving, but they get rapidly dissolved by the hot shock gas flowing past them \citep[see also][]{bieri17,richings17}.

Our analytical model focuses on isolated galaxies to better study the influence of individual parameters, independently from the cosmological environment. Therefore we are not able to take into account large-scale effects, such as the clustering of galaxies \citep{gaspari11,gaspari11b,gaspari14}, the accretion of cold gas along cosmic filaments, or galaxy mergers.

We assume the AGN luminosity to be constant during the lifetime of the AGN. This is a necessary assumption to clearly disentangle the influence of different input parameters. Different groups study the radiation-regulated accretion on to the BH and demonstrate the importance of multidimensional hydrodynamical simulation to self-consistently follow the accretion and resulting luminosity \citep{park11,sugimura16,negri17}. The accretion rate and hence the AGN luminosity oscillate by up to two orders of magnitude on time-scales of several thousand years, shorter than the lifetime of the AGN in our model.
\cite{gilli17} investigate galactic outflows driven by an AGN with exponentially increasing luminosity, i.e. they self-consistently account for the mass growth of the central BH at constant Eddington ratio. They find that the late time expansion of the radius in their 1D model is also exponential, irrespective of the details of the driving mechanism.

We only distinguish between the momentum- and energy-driven phase, whereas \citet{FG12} also account for the ``intermediate partially radiative bubble stage''. In this phase, the cooling time is shorter than the flow time but longer than the crossing time of the shock and the shock cools only partially. Although we agree that a more realistic transition from one driving mechanism to another is necessary, this does not affect our fiducial model since the transition occurs already very early and is very sharp.

The weak collisional coupling between protons and electrons in the shocked wind can increase the Compton cooling time by two orders of magnitude (Sec. \ref{sec:twot}). We do not implicitly include this effect, but as pointed out by \citet{FG12}, the more realistic treatment of the shock as a two-temperature medium significantly decreases the efficiency of inverse Compton cooling. This does not change anything in our fiducial model, because the transition to energy-driven occurs already at very small radii. However, the AGN transition luminosity, above which the outflow is always momentum-driven in the disc (Eq. \ref{eq:MBHtrans}) will be higher as it scales $L_\mathrm{AGN,crit}\propto t_\mathrm{compton}^2$.

An additional increase of the Compton cooling time can be achieved by the inclusion of Compton heating. \citet{sazonov04} determine the equilibrium Compton temperature of gas exposed to a characteristic AGN spectrum to be $T_C=2 \times 10^7$\,K, which is insensitive to obscuration effects. The equilibrium Compton temperature is equal to the mean photon energy averaged over the AGN spectra. They conclude that the characteristic Compton heating and cooling rates per particle should be the same within a factor of $\sim 2$. Most of the Compton cooling will be provided by the IR component, whereas the Compton heating is dominated by the high-energy component. The maximum radius out to which an AGN can heat low density, highly ionized gas by Compton heating is
\begin{equation}
r_C = 0.4\,\mathrm{kpc}\left( \frac{f_\mathrm{Edd}}{0.1} \right) ^{1/2} \left( \frac{M_\mathrm{BH}}{10^8 \Msun} \right) ^{1/2}.
\end{equation}
For ionization parameters below $\xi \lesssim 10^5$ the gas cannot be heated to $T_C$, due to other more efficient cooling mechanisms. A more detailed discussion of the effect of Compton heating is given in \cite{sazonov04,sazonov05}.

Hydrodynamical simulations of a geometrically thin and optically thick accretion disc show that the disc wind might not be isotropic, as assumed in our model, but has a covering factor of $\sim 20\%$ \citep{proga00}. Moreover, the initial wind velocity itself might depend on the angle to the disc with higher velocities in the perpendicular direction and a possible self-shadowing effect by the outer part of the disc amplifies this anisotropy \citep{sugimura16}.



\section{Summary and Conclusion}
We present a new 2D analytical model for AGN-driven outflows and demonstrate the importance of a more realistic gaseous disc profile for the outflow dynamics. In contrast to simplistic 1D models, we predict smaller gas ejection fractions and shorter fallback times, both facilitating an efficient mass growth of BHs via feeding from galactic scales. This is related to the energy-driven nature of the outflows, which pave a path of least resistance perpendicular to the disc and hence prevent efficient driving in the disc plane. We confirm earlier results that AGN-driven winds can transport energy and momentum from sub-pc scales around the BH to galactic scales and thereby regulate the co-evolution of a galaxy and its central BH.

Our main results are as follows:
\begin{itemize}

\item For typical high-redshift galaxies, the AGN can eject at most $\sim 10\%$ of the ISM out of the halo, whereas 1D models predict complete gas ejection under similar conditions.  At high redshift, the ejected gas mass fractions are lower due to the deeper gravitational potential, compared to similar galaxies at low redshift.

\item The characteristic time for which the AGN can suppress further gas supply is remarkably short, of the order a few million years. We find AGNs with a low luminosity to be more efficient in converting their input energy into outflow momentum, because the swept-up medium is confined for longer by the shock front and has consequently more time to store up internal energy.

\item We also find a systematic transition in the outflow nature: for AGN luminosities below $10^{43}\,\mathrm{erg}\,\mathrm{s}^{-1}$ the outflow is energy-driven, independent of the halo properties. At higher luminosities, the outflow remains momentum-driven in the disc plane.

\item Independently of the exact criterion to suppress further gas accretion, we find a slope of $M_\mathrm{BH} \propto \sigma ^{4.8}$ for the $M$--$\sigma$ relation.
\end{itemize}

Our new model highlights the importance of a realistic 2D density profile to predict the ejected gas mass fraction and fallback time. We can reproduce results of 3D hydrodynamical simulations with much less computational effort, although more realistic models are still necessary to correctly account for the substructure of the ISM and for environmental effects. Our results can be used as an improved subgrid model in cosmological simulations or semi-analytical models of galaxy formation to predict the efficiency of AGN feedback.


\subsection*{Acknowledgements}
We thank the reviewer for her/his suggestions and careful reading of the manuscript.
We are grateful to Alex Wagner and Rebekka Bieri for valuable discussions and helpful comments on our model.
The authors acknowledge funding under the European Community's Seventh Framework Programme (FP7/2007-2013) via the European Research Council Grants `BLACK' under the project number 614199.

\bibliographystyle{mn2e}
\bibliography{2DAGNoutflows}

\bsp
\label{lastpage}

\end{document}